\documentclass[twocolumn]{aastex631} 
\usepackage{amsmath}
\usepackage{mathtools}
\usepackage{makecell}
\usepackage{pdflscape}
\makeatletter
\AtBeginDocument{\let\LS@rot\@undefined}
\makeatother
\graphicspath{{./}{figures/}}
\submitjournal{PSJ}
\shorttitle{Abundance Measurements of Interstellar Comets}
\shortauthors{Seligman et al.}
\begin{document}
\title{The  Volatile Carbon to Oxygen Ratio as a Tracer for the Formation Locations of Interstellar Comets}

\author[0000-0002-0726-6480]{Darryl Z. Seligman}
\affiliation{Dept. of the Geophysical Sciences, University of Chicago, Chicago, IL 60637}

\correspondingauthor{Darryl Seligman}
\email{dzseligman@uchicago.edu}

\author[0000-0003-0638-3455]{Leslie~A.~Rogers}
\affiliation{Department of Astronomy and Astrophysics, University of Chicago, Chicago, IL 60637, USA}

\author[0000-0001-9749-6150]{Samuel H. C. Cabot}
\affil{Dept. of Astronomy, Yale University, 52 Hillhouse, New Haven, CT 06511, USA}

\author[0000-0003-2152-6987]{John W. Noonan}
\affil{Department of Physics, Auburn University, AL, USA}

\author[0000-0002-1422-4430]{Theodore Kareta}
\affil{Lowell Observatory, Flagstaff, AZ, USA}

\author[0000-0001-8397-3315]{Kathleen E. Mandt}
\affiliation{Johns Hopkins Applied Physics Laboratory, Laurel, MD, USA}

\author[0000-0002-0093-065X]{Fred Ciesla}
\affiliation{Dept. of the Geophysical Sciences, University of Chicago, Chicago, IL 60637}
\author[0000-0002-0622-2400]{Adam McKay}
\affiliation{American University/NASA Goddard Space Flight Center, Greenbelt, MD, 20771}

\author[0000-0002-9464-8101]{Adina~D.~Feinstein}
\altaffiliation{NSF Graduate Research Fellow}
  \affiliation{Department of Astronomy and Astrophysics, University of Chicago, Chicago, IL 60637, USA}
  \author[0000-0002-1422-4430]{W. Garrett Levine}
\affil{Dept. of Astronomy, Yale University, 52 Hillhouse, New Haven, CT 06511, USA}

\author[0000-0003-4733-6532]{Jacob L. Bean}
\affiliation{Department of Astronomy and Astrophysics, University of Chicago, Chicago, IL 60637, USA}

  \author[0000-0001-5344-8069]{Thomas Nordlander}
  \affiliation{Research School of Astronomy and Astrophysics, Australian National University, Canberra, ACT 2611, Australia}
\affiliation{ARC Centre of Excellence for Astronomy in Three Dimensions (ASTRO-3D), Canberra, ACT 2611, Australia}

\author[0000-0003-3893-854X]{Mark R. Krumholz}
\affiliation{Research School of Astronomy and Astrophysics, Australian National University, Canberra, ACT 2611, Australia}
\affiliation{ARC Centre of Excellence for Astronomy in Three Dimensions (ASTRO-3D), Canberra, ACT 2611, Australia}

\author[0000-0003-4241-7413]{Megan Mansfield}
\affiliation{Steward Observatory, University of Arizona, 933 North Cherry Avenue, Tucson, AZ 85721, USA}

\author[0000-0002-7783-6397]{Devin J. Hoover}
\affil{Dept. of Astronomy and Astrophysics, University of Chicago, Chicago, IL 60637}

\author[0000-0002-5954-6302]{Eric Van Clepper}
\affiliation{Dept. of the Geophysical Sciences, University of Chicago, Chicago, IL 60637}

\begin{abstract}
Based on the occurrence rates implied by the discoveries of 1I/`Oumuamua and 2I/Borisov, the forthcoming Rubin Observatory Legacy Survey of Space and Time (LSST)  should detect $\ge1$ interstellar objects  every year \citep{Hoover2021}. We advocate  for future  measurements of the production rates of  H$_2$O, CO$_2$ and CO in these objects to estimate their carbon to oxygen ratios, which traces formation locations within their original protoplanetary disks. We review similar measurements for Solar System comets, which  indicate formation interior to the CO snowline. By quantifying the relative processing in the interstellar medium  and  Solar System, we estimate  that production rates will not be representative of  primordial compositions for the majority of interstellar comets. Preferential desorption of CO and CO$_2$ relative to H$_2$O in the interstellar medium implies that measured C/O ratios represent lower limits on the primordial ratios. Specifically,  production rate ratios of ${\rm Q}({\rm CO})/{\rm Q}({\rm H_2O})<.2$ and ${\rm Q}({\rm CO})/{\rm Q}({\rm H_2O})>1$ likely indicate formation interior and exterior to the CO snowline, respectively.  The  high C/O ratio of 2I/Borisov implies that it formed exterior to the CO snowline. We provide an overview  of  the currently operational facilities capable of obtaining these measurements that will  constrain the fraction of ejected comets that formed exterior to the CO snowline. This fraction will provide key insights into   the efficiency of and mechanisms for cometary ejection in exoplanetary systems.  %Finally, we discuss  additional scientific benefits from obtaining in situ measurements  of an interstellar comet.
\end{abstract}
\section{Introduction} \label{sec:intro}

The composition and  activity of comets, and how this reflects  their primordial composition, formation location and dynamical evolution has  been a long-standing subject of inquiry.  In 1812, William Herschel obtained detailed observations of two comets, both of which attained different brightness levels despite similar perihelia distances \citep{Herschel1812, Herschel1812b}. In order to account for this discrepancy,  he speculated that the brighter comet originated from interstellar space and acquired ``unperihelioned matter by moving in a parabolical direction through the immensity of space.'' Laplace contemporaneously performed a surprisingly accurate estimate for the number of interstellar comets that should pass close to the Sun's vicinity  \citep{Laplace1814,Heidarzadeh2008}.

The current, larger census of  comets  can  be sorted into two populations: ecliptic comets, a subset of which are Jupiter Family Comets (JFCs), and Long Period Comets (LPCs) which have isotropic distributions of inclination. It is generally believed  that the JFCs originate in  the trans-Neptunian region  \citep{Leonard1930,Edgeworth1943,Edgeworth1949,Kuiper1951,Cameron1962,Whipple1964,Everhart1972,Vaghi1973,Joss1973,Delsemme1973,Fernandex1980MNRAS,Duncan1988,Quinn1990,Jewitt1993,Prialnik2020}, and migrate  into the inner Solar System via the Centaur region  \citep{Hahn1990,Levison1997,Tiscareno2003,DiSisto2007,Bailey2009,DiSisto2009,Nesvoryn2017,Fernandez2018,Sarid2019,SeligmanKratter2021}, while the LPCs come from  the Oort cloud \citep{Oort1950}. 

While the dynamical evolution of the trajectory of a comet reveals its recent whereabouts,  compositional measurements can trace the original formation location -- in the absence of significant post-formation processing.  Specifically, the ratio of combinations of molecular  production rates give  elemental abundance ratios. There is a precedent for using the carbon to oxygen (C/O) ratio as a tracer of the formation location of giant exoplanets within their protostellar disk, as this ratio should increase between the H$_2$O, CO$_2$ and CO freezeout snowlines \citep{Oberg2011}. For a Solar System cometary analogue to these exoplanetary measurements, it was generally believed that  the  JFCs formed exterior to the CO snowline, while the LPCs formed between  the giant planets,  and that their compositions  would provide evidence for this (see \citet{Rickman2010} and references therein for a recent review). However,  space based spectroscopic measurements of CO$_2$ and CO production rates \citep{Ootsubo2012}  revealed a surprisingly low C/O ratio in almost all Solar System comets, implying formation interior to the CO snowline \citep{Ahearn2012}. Ground-based infrared spectroscopy has demonstrated that the JFCs tend to be depleted in CO compared to the LPCs, although the sample size is small due to the difficulty of these measurements \citep{DelloRusso2016,Disanti2017,Roth2018,Faggi2019,Roth2020,McKay2021}. A recent compositional survey of  CO, CO$_2$, and H$_2$O for 20 cometary objects has confirmed this result \citep{Pinto2021} and (Harrington Pinto et al., Under Review, 2022).  CO  activity is  observed in distant Centaurs \citep{Senay1994,Crovisier1995,Womack1997,Womack1999b,Choi2006,Bauer2008,Gunnarsson2008,Jewitt2009,Jaeger2011,Paganini2013,Bauer2015,Womack2017,Wierzchos2017,Schambeau2018,James2018,Kareta2019,Wierzchos2020}, but diminished levels of H$_2$O activity due to low ambient temperatures prohibits accurate measurements of the  volatile C/O ratio in these objects.

Given the efficacy with which the Solar System ejected planetesimals via planetary migration and/or instability \citep{Hahn1999,Gomes2004,Tsiganis2005,Morbidelli2005,Levison2008,Raymond2018b,Raymond2020}, it is feasible that  CO-enriched comets  formed exterior to the CO snowline,  most of which were ejected into interstellar space. An  intriguing object is C/2016 R2 \citep{Wierzchos2018, Cochran2018,McKay2019}, an almost hyperbolic LPC with a CO production rate  $\gtrsim100$ that of H$_2$O \citep{McKay2019}, suggestive that it formed exterior to the CO snowline. 

The exotic composition of R2  led \citet{McKay2019} to speculate that it formed outside of our Solar System. Astronomers had considered the presence of interstellar comets in the Solar System prior to the detection of R2.  The number density of interstellar comets was predicted based on non-detections \citep{sekanina1976probability, mcglynn1989nondetection,francis2005demographics,Moro2009, Engelhardt2014} with all sky surveys such as  Pan-STARRS \citep{jewitt2003project,Chambers2016}.  The forthcoming LSST \citep{jones2009lsst,Ivezic2019}, whose ability to detect transient objects has been demonstrated \citep{solontoi2011comet,Veres2017,veres2017b,Jones2018},  was projected to detect  between 0.001 and 10 interstellar comets inferred from early estimates derived from non-detections \citep{Cook2016}. 

In 2017, well before LSST's first light, the first interstellar object  1I/2017 U1 (`Oumuamua) was discovered. However, observations obtained in order to measure the volatile production rates produced non-detections \citep{Meech2017,Jewitt2017,trilling2018spitzer}. The object exhibited an extreme $6:6:1$ geometry \citep{Knight2017,Bolin2017,Fraser2017,McNeill2018,Belton2018,Mashchenko2019,Drahus2017}, a non-gravitational acceleration \citep{Micheli2018}, and a moderately reddened  color \citep{masiero2017spectrum,Fitzsimmons2017,Bannister2017,Ye2017} consistent with its young $<40$~Myr age \citep{mamajek2017,Gaidos2017a, Feng2018,Fernandes2018,hallatt2020dynamics,Hsieh2021}. 

This peculiar combination of unique physical properties  led to a variety of theories regarding the provenance of `Oumuamua. If the non-gravitational acceleration was  driven by radiation pressure \citep{Micheli2018}, this  would imply that `Oumuamua was an ultra low-density fractal aggregate \citep{moro2019fractal,luu2020oumuamua,Sekanina2019b,Flekkoy2019}, or an artificial millimeter thin membrane \citep{bialy2018radiation}. An artificial origin could not be confirmed, since no radio signals were found to be coming from the object \citep{Enriquez2018,Tingay2018,Harp2019}. If the acceleration was powered by outgassing \citep{seligman2019acceleration}, the  energetics could be consistent with a bulk composition of H$_2$ \citep{fuglistaler2018solid,seligman2020,Levine2021_h2}, N$_2$ \citep{jackson20211i,desch20211i}, or CO \citep{Seligman2021}. Other theories invoke a tidally fragmented planetesimal \citep{Raymond2018b,zhang2020tidal}, and ejection from a post-main sequence star system \citep{Hansen2017,Rafikov2018b,Katz2018}, or circumbinary system \citep{Cuk2017,Jackson2017}.  However, the anomalous acceleration  largely ruled-out these interpretations.  \citet{Flekkoy2022} calculated observable spectral signatures that will differentiate between  proposed formation theories in future objects. 

A second interstellar object, 2I/Borisov, was detected in 2019. This confirmed  the existence of a galactic population of interstellar objects with spatial number densities of order $n_{o}\sim1-2\times 10^{-1}\,$\textsc{au}$^{-3}$ \citep{Trilling2017,Laughlin2017,Jewitt2017,Rafikov2018,Zwart2018,Do2018,moro2019a,moro2018,Levine2021}.  2I exhibited  a  dusty coma \citep{Jewitt2019b,Bolin2019,Fitzsimmons:2019,Ye:2019,McKay2020,Guzik:2020,Hui2020,Kim2020,Cremonese2020,yang2021} with  typical cometary carbon- and nitrogen-bearing species detected \citep{Opitom:2019-borisov, Kareta:2019, lin2020,Bannister2020,Xing2020,Aravind2021}.  2I  was enriched in CO relative to H$_2$O \citep{Bodewits2020, Cordiner2020}, indicating formation exterior to the CO snowline in its original protoplanetary disk \citep{Price2021}. \citet{Lisse2022} argued that the CO enrichment could be explained if it  was ejected within $<20$MYRs of the formation of  its host system. It exhibited an outburst  \citep{Drahus2020ATel} and  fragmentation event \citep{Jewitt2020:BorisovBreakup, Jewitt2020ATel, Bolin2020ATel,Zhang2020ATel} due to seasonal effects \citep{Kim2020}. Its non-gravitational acceleration was consistent with measured production rates \citep{Hui2020,delafuente2020,Manzini2020}. Curiously, atomic nickel vapor \citep{Guzik2021} and abnormally high polarization  \citep{Bagnulo2021} were detected in the coma. 

It is feasible that Borisov --- and even possibly `Oumuamua if the CO hypothesis is correct --- are representative of CO-enriched comets that formed exterior to the CO snowline. Moreover, R2 may be one such comet that survived ejection from our own Solar System; although, its considerably higher  CO/H$_2$O ratio than that of Borisov and  unexpectedly high N$_2$ abundance are difficult to explain via typical formation mechanisms \citep{Wierzchos2018,Mousis2021}. Future detections and compositional measurements will demonstrate whether or not these  objects are representative of  the  population.  The LSST should detect $\ge1$ interstellar object every year \citep{Hoover2021} and  a space based in situ rendezvous may be performed in the upcoming decade \citep{Seligman2018,Hein2017,Meech2019whitepaper,Castillo-Rogez2019,Hibberd2020,Donitz2021,Meech2021,Hibberd2022,Moore2021whitepaper}.  ESA’s  \textit{Comet Interceptor} \citep{jones2019,Sanchez2021} mission or the NASA concept study \textit{BRIDGE} \citep{Moore2021} may provide these observations. 

In this paper we demonstrate that  the volatile C/O ratio of interstellar comets encodes their formation location with respect to the CO snowline, and outline how this ratio can be measured with currently operational facilities. This paper is organized as follows: In \S \ref{sec:CO_tracer}, we demonstrate that the volatile C/O ratio will serve as a tracer of formation location in an interstellar comet. In \S \ref{sec:stellar_CO}, we review measured stellar abundance ratios, as well as their systematic uncertainties and dependencies on metallicity. In \S \ref{sec:cometary_CO}, we review measured abundances of Solar System comets, and establish CO$_2$, CO and H$_2$O as good tracers for the C/O ratio of a comet. We show that most Solar System objects with  compositions measured likely formed interior to the CO snowline, while 2I/Borisov and C/2016 R2 likely formed exterior to it. In \S\ref{sec:processing}, we quantify  the extent to which measurements of the coma of an interstellar comet are representative of the primordial composition, based on processing in the interstellar medium and  Solar System. We  show that the majority of detected interstellar comets will not exhibit activity representative of their primordial compositions. Therefore, measured C/O ratios are lower limits due to the preferential desorption of CO and CO$_2$ with respect to H$_2$O in the interstellar medium.  In \S \ref{sec:predictions}, we identify  C/O ratios that are definitive indications of formation exterior and interior to the CO snowline, \textit{after} the ISM processing.   In \S \ref{sec:observations}, we outline observational facilities that can measure production rates of H$_2$O, CO and CO$_2$ in an interstellar comet, and conclude in \S \ref{sec:conclusions}.

\begin{figure}
\begin{center}
\includegraphics[width=\linewidth]{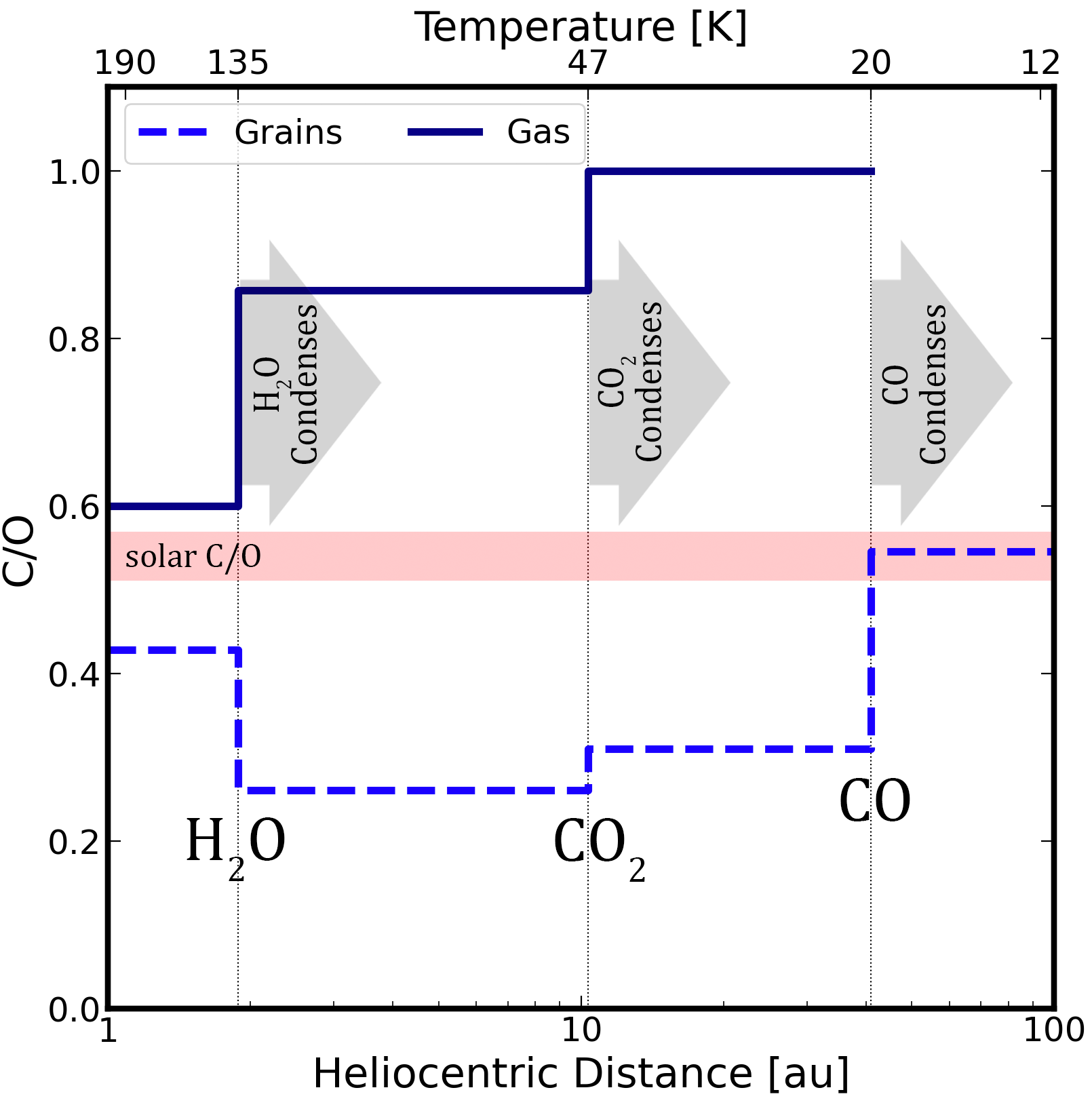}
\caption{The C/O ratio of grains and gas as a function of heliocentric distance. The ratio is impacted by condensation of oxygen- and carbon-bearing molecules at and beyond their respective snowlines. The C/O profile of gas ends at the CO snowline because all major C- and O-bearing species haved condensed beyond this point. Comets that form exterior to the CO snowline have higher C/O ratios than those that form interior to it. This figure is a rendition of Figure 1 in \citet{Oberg2011}.}\label{Fig:c_o_distance}
\end{center}
\end{figure}

\section{The C/O Ratio Traces Comet Formation Location}\label{sec:CO_tracer}

As presented by \citet{Oberg2011} and references therein, H$_2$O, CO$_2$ and CO are abundant oxygen-bearing and carbon-bearing molecules in protoplanetary disks, which motivates consideration of carbon and oxygen as tracers of formation conditions. Importantly, these molecules condense at radii where planets are expected to form, and where large masses of solids are detected in protoplanetary disks (Figure~\ref{Fig:c_o_distance}). Observational evidence of such ``snowlines" may be found based on the spatial distribution of molecular or ion tracers, or through changes in the dust size distributions in the disk \citep{Qi2013,Banzatti2015}. The relative amounts of H$_2$O, CO$_2$ and CO in gaseous versus solid phase will alter the C/O ratio of minor bodies. The C/O of grains, assumed to contain some refractory carbon, decreases exterior to the H$_2$O snowline, since H$_2$O condenses and contributes oxygen to solid materials. The ratio then rises outside of the CO$_2$ and CO snowlines due to the subsequent incorporation of these species in icy material (Figure~\ref{Fig:c_o_distance}). The most dramatic differences in the bulk composition and C/O ratio specifically are for objects that form  interior and exterior to the CO snowline. This also defines a natural demarcation between the ``outer'' and ``inner'' region of the system, since the CO snowline is currently  located approximately at Neptune's distance in our own Solar System. It is important to note that the locations of these snowlines migrate as the star and disk evolve \citep[e.g.][]{cieza16}.

Before focusing our attention on comets, we briefly point out the extensive precedent for using the C/O ratio to trace the formation location and evolution of extrasolar planets. Theoretical predictions indicate that carbon-rich environments can produce solids such as SiC, TiC and graphite as opposed to silicate-based building blocks in our own Solar System \citep{Bond2010}. Similarly, carbon and oxygen are important constituents in hydrogen-dominated, gas giant atmospheres \citep{Burrows1999, Fortney2010, Moses2011}. Whether the C/O ratio is greater or less than unity heavily influences the production of certain molecular species, including H$_2$O, HCN, C$_2$H, and CH$_4$ \citep{Seager2005, Madhusudhan2012}. In turn, the C/O ratio of planetary interiors and atmospheres can be used to infer formation conditions and approximate locations within the host protoplanetary disk \citep{Oberg2011}. These considerations have been crucial for testing whether hot Jupiters ($P < 10$~days) formed {\it in situ} or experienced inward migration. We refer the reader to  \citet{Dawson2018} for a comprehensive review of formation theories. 

The H$_2$O, CO$_2$, and CO abundances can also  be used to trace formation conditions of comets, which represent  planetesimals that formed exterior to the H$_2$O snowline. \citet{Ahearn2012} attributed the scatter in production rate ratios of CO/H$_2$O and of CO$_2$/H$_2$O of comets to their primordial compositions, as opposed to chemical evolution from consecutive perihelion passages. They concluded that the majority of comets formed between the CO$_2$ and CO snowlines, which also explains the lower degree of scatter in CO/CO$_2$ measurements.  In  contrast to most of the Solar System comets, 2I/Borisov's CO/H$_2$O was greater than unity and likely formed   exterior to the CO snowline in its original protoplanetary disk \citep{Bodewits2020,Cordiner2020}. %In the following section, we discuss the effects of stellar C/O ratio in cometary composition. 

\section{Measurements of  Stellar Carbon to Oxygen  Ratios}\label{sec:stellar_CO}
In this section, we describe our current understanding of the C/O ratios in stars that produce interstellar comets. We discuss systematic uncertainties for measured stellar abundance ratios  and  the current ability to compare  these measurements  to cometary compositional measurements.  This is essential for inferring the formation location of an interstellar comet  within the protostellar disk of its host star. 

\subsection{Carbon to Oxygen Ratios in Solar Twins}
A fundamental complication regarding measuring and comparing C/O ratios for different stars is that the stellar variations in this ratio are relatively small. Therefore, accurately comparing stellar C/O ratios  requires a high level of both precision and accuracy. In stars with similar surface temperature and gravity to the Sun, commonly referred to as solar twins or solar type stars, the C/O ratio can be measured with a high precision. The calibration to the Sun implies that the systematic errors of abundance ratios for solar twins are almost the same for each star and mostly cancel. 

\begin{figure}
\begin{center}
\includegraphics[scale=0.45,angle=0]{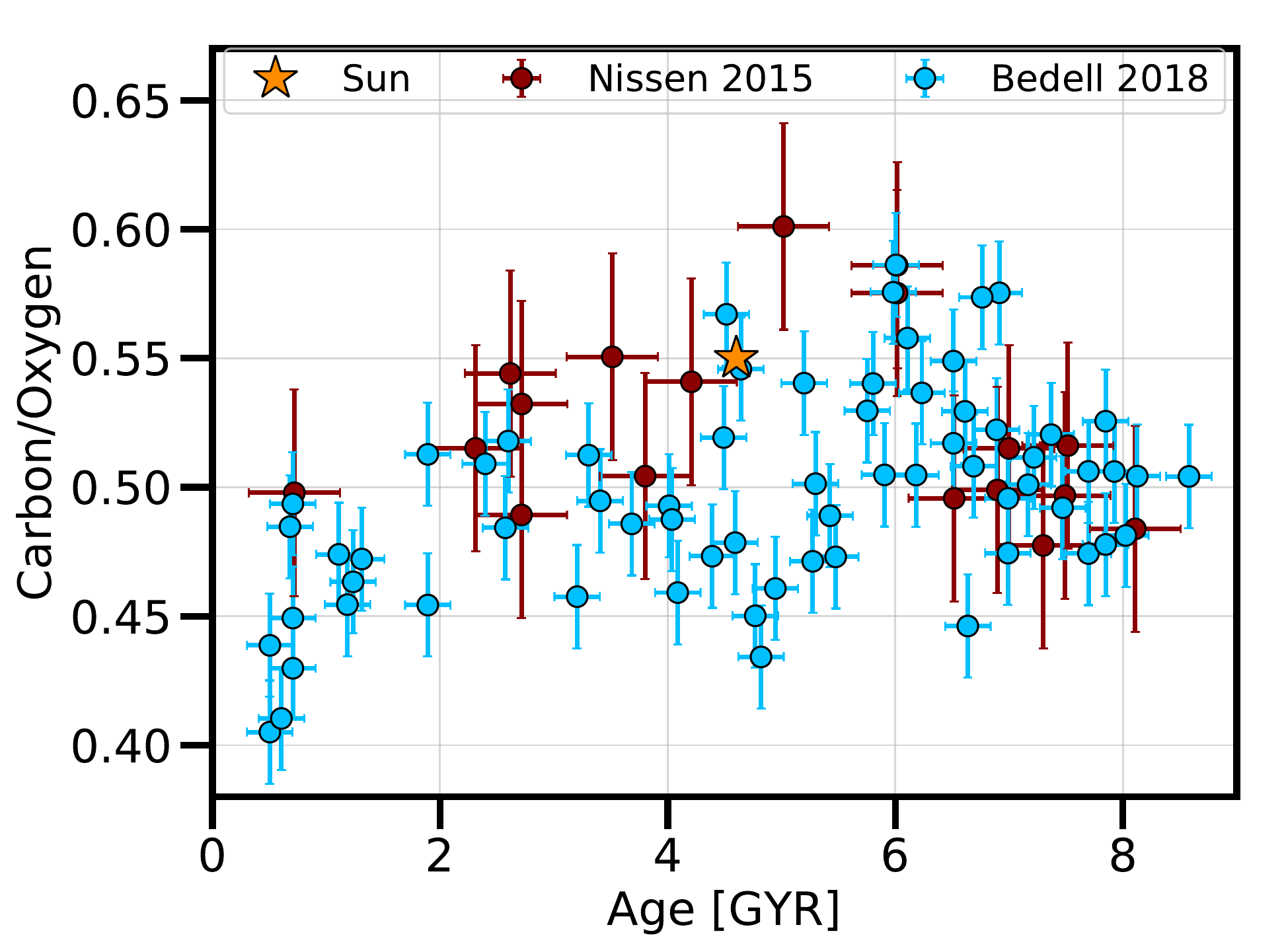}
\caption{Measured C/O ratios in solar twins from \citet{Bedell2018} and \citet{Nissen2015}. The resulting combined data set has a mean C/O ratio of $\sim 0.50$ with a standard deviation of $\sigma\sim$0.04.}\label{Fig:twins}
\end{center}
\end{figure}

\citet{Bedell2018} performed a  differential study of the chemical composition of solar twins, and showed that the C/O ratio in the sample did not vary significantly. However, they  reported a slight dependence of this ratio on the stellar metallicity, which is a measure of heavy-element enrichment resulting from galactic chemical evolution. \citet{Nissen2015} presented an analogous analysis for a sample of 21 solar twins, and reported similar results.  The  reported C/O ratio for the solar twins measured by both studies are shown in Figure \ref{Fig:twins} (with data drawn from Figure 11 of \citet{Nissen2015} and Figure 7  of \citet{Bedell2018}). While the lowest metallicity stars have lower C/O ratios, the mean C/O ratio of all of these stars is 0.50, and the standard deviation is $\sigma=0.04$. All of these measurements are calibrated to measured solar abundances, which can be measured with very low uncertainties using a 3-dimensional local thermodynamic equilibrium (LTE) analysis of molecular lines  \citep{Amarsi2021}.

\subsection{Carbon to Oxygen Ratios in All Stars}
Extrapolating the C/O ratio for stars with different masses and temperatures than the Sun is less straightforward. \citet{Brewer2016} and \citet{Delgado2021} performed almost-differential studies of stars that are similar but not identical to the Sun. These authors reported high precision for samples of 1,617 \citep{Brewer2016} and 1,111 F, G, and K stars \citep{Delgado2021}, but systematic errors in the measurements are present. Figure 5 in \citet{Delgado2021} shows that the C/O ratio does vary with age, similar to what was found by \citet{Nissen2015} and \citet{Bedell2018}.

The largest source of systematic uncertainty in stellar C/O measurements is that the inferred abundance ratios are derived from spectral lines  that  do not form under LTE, and   require  advanced radiative transfer calculations. Ultimately, these discrepancies vary primarily with  $T_\text{eff}$, $\log(g)$, and [Fe/H] \citep{Amarsi2019}. Encouragingly, the differences are  small for stars where $|T_\text{eff}-T_\odot|\le$ 100 K, so solar twin measurements outlined in the previous subsection that do not incorporate non-LTE effects are still reliable.

An example of how these systematic errors work is shown in Figure 7 of \citet{Bedell2018},   which compares their  data on solar twins to the more general sample of \citet{Brewer2016}.  \citet{Bedell2018}  demonstrated that the true  C/O ratio  varies slightly from star to star with an effect that is several times larger than their measurement errors; the standard deviation of the sample is $\sim0.04$, while the estimated measurement uncertainties are $\sim0.02$. These results are also very similar to those of the advanced non-LTE study by \citet{Amarsi2019}, even though this sample included stars with a much wider range of stellar parameters.

The  analysis presented by \citet{Amarsi2021}   indicated that the limiting factor in measuring the solar C/O ratio  is the atomic and molecular data, rather than the model solar atmosphere. Therefore, errors on the absolute abundances of carbon and oxygen do not cancel, but should be combined in quadrature to calculate the uncertainty on the solar C/O ratio.  The quoted accuracy is  $\sim0.05$ dex, where $\log (N_\text{O} / N_\text{C}) = 0.23 \pm 0.05$\,dex, or C/O = 0.59 $\pm$ 0.06. This uncertainty only applies to the absolute abundance scale. Therefore, for any  homogeneous population of stars (like Solar twins), the uncertainty in the scatter of the distribution  is much less than 0.05 dex. 

For stars that are very unlike the Sun, such as cool M-type stars, the spectra are veiled by millions of molecular lines. These introduce a different set of systematic errors which have still not been quantified and solved. \citet{Veyette2016} indicated that while the C/O ratio  influences M-dwarf spectra, it is unclear if synthetic spectra are good enough to fit this to a precision better than the star-to-star scatter. In case  measurements  of the C/O ratio in M-dwarfs are acquired and calibrated on FGK stars, the 0.05 dex accuracy of the solar C/O ratio will limit those results as well. Increasing the accuracy of the atomic and molecular constants via improved laboratory measurements and theoretical calculations is the only way to reduce this dominant source of uncertainty.

\subsection{Assumptions For Interstellar Comet Calculations}

 The population of interstellar comets originate from an unknown and undifferentiated assortment of  stellar populations. Therefore, in this study, we assume that the bulk composition and mean and scatter of elemental ratios of these stars are similar to the solar twins.  Our justification for this is that there is no process  that would differentially change the abundance ratios of interstellar comet producing stars as a function of stellar mass. It is  important to note that the location of the true mean has 0.05 dex uncertainty. Therefore the predictions presented in Section \ref{sec:predictions} would scale based on the true mean, when it is further constrained.
\section{Measurements of Cometary C/O Ratios}\label{sec:cometary_CO}

A classical picture of cometary formation is that the JFCs formed in the current trans-Neptunian region, while  the LPCs formed between  the giant planets. \citet{Ahearn2012} presented an overview of abundance measurements in a sample of comets that had measured production rates of CO, CO$_2$ and H$_2$O. Based on the relatively low abundance of CO in these comets, they argued that all comets formed in the giant planet region between the CO and CO$_2$ snowlines, contrary to the classical picture. Their measurements demonstrated  that JFCs  formed slightly closer to the Sun than  the LPCs.  Recent reviews of the observed compositional properties of comets can be found in \citet{Rickman2010}, \citet{Cochran2015}, \citet{Biver2016} and \citet{Bockelee2017}.

While comets exhibit a wide variety of compositions,  H$_2$O is the primary driver of activity in the inner Solar System for most of the population.  CO and CO$_2$ are the most common volatiles besides H$_2$O (Table 1 and Figure 2 in \citet{Bockelee2017}). Typical comets   consist of $0.2-23\%$ CO and  $2.5-30\%$ CO$_2$ in their coma compared to H$_2$O  - where percentages are calculated by number of molecules. CHO molecules and hydrocarbons make up on average  $\sim4$ and $\sim2\%$, while nitrogenous molecules and sulfur-bearing molecules constitute $\sim1.5$ and $\sim 1 \%$. The most common species in comets on average after CO and CO$_2$ is CH$_3$OH, followed by C$_2$H$_6$, CH$_4$ and H$_2$S, but these are typically at least an order of magnitude less abundant than CO$_2$, and at least  factor of a few less abundant than CO. Therefore, only measuring CO, CO$_2$, and H$_2$O in a cometary coma can provide a reasonable first  approximation of the volatile C/O ratio.

\begin{figure}
\begin{center}
\includegraphics[scale=0.45,angle=0]{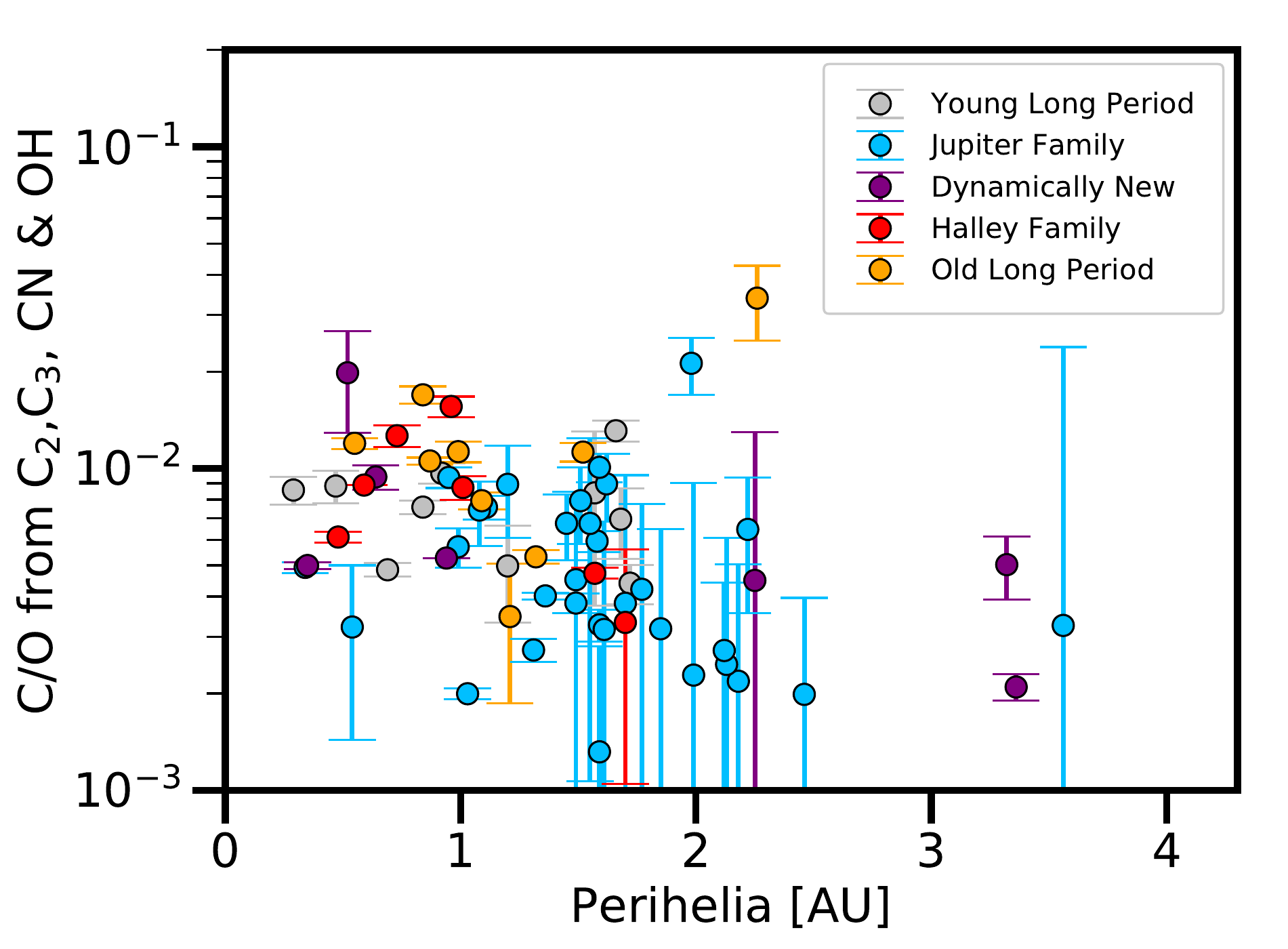}
\caption{The C/O ratios for comets calculated using weighted production rates of C$_2$, C$_3$, CN to H$_2$O (from OH) production rates. These comets are in the sample presented by \citet{Ahearn:1995}. The error bars are calculated by adding reported uncertainties in quadrature scaled by the weighting of the species for the abundance ratio. }
\label{fig:ahearn_comets}
\end{center}
\end{figure}
\subsection{Carbon and Oxygen Measured in Cometary Bodies}
 In this subsection we review currently measured compositional properties of Solar System comets, and calculate the inferred C/O ratios using a variety of combinations of species. As a starting place, we refer to the sample of 87 comets with measured production rates of carbon bearing  molecules C$_2$, C$_3$ and CN, and OH, over a period of 17 years \citep{Ahearn:1995}. These comets do not have measured production rates of CO and CO$_2$. In Figure \ref{fig:ahearn_comets}, we show the C/O ratio of these comets calculated using \textit{only} the production rates of C$_2$, C$_3$, and CN relative to  H$_2$O (inferred from OH). The conversion from OH to H$_2$O is calculated using  the branching ratio of H$_2$O photodissociation and the heliocentric distance \citep{Cochran1993,McKay2019}. These objects span a range of cometary families, and  every estimated C/O ratio is $<0.1$ --  less than that expected  between the H$_2$O and CO$_2$ snowlines (Figure~\ref{Fig:c_o_distance}). The C/O ratio tends to increase with decreasing heliocentric distance due to the increase in C$_2$/CN ratios, which is also reflected when plotted as a function of perihelia (see Fig. 3, panel A and Figure 15 of \citet{Ahearn:1995} and \citet{Fink2009}).  This is likely related to the fact that, in addition to being released via photodissociation of more complex molecules, these species are  also released from thermal degradation of carbonaceous dust grains -- such as the CHON grains discovered during the Halley flyby \citep{Lawler1992} -- a process that is more efficient closer to the Sun.  

\begin{figure*}
\begin{center}
\includegraphics[scale=0.6,angle=0]{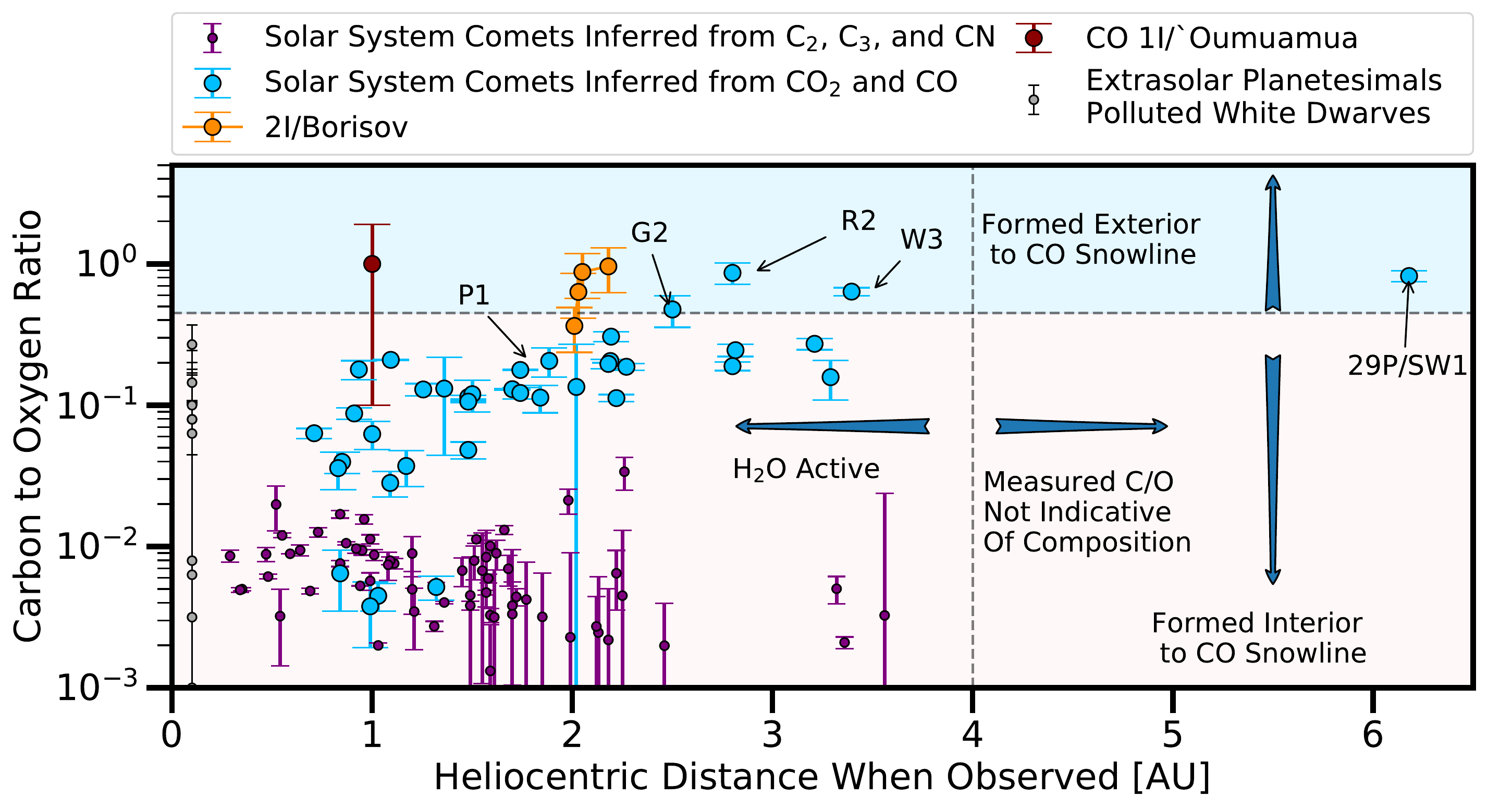}
\caption{An overview of measurements of the carbon to oxygen ratio in  Solar System  and interstellar comets. The comets presented in \citet{Ahearn:1995} with only C$_2$, C$_3$, CN to OH production rates measured are shown in purple points. The comets with measured CO$_2$, CO and H$_2$O production rates, listed in Table \ref{tab:production rates}, are shown in blue points. The C/O ratio of Borisov and `Oumuamua are shown in gold and red, assuming that the acceleration of `Oumuamua was caused by CO outgassing. We include points for all 4 CO detections of 2I/Borisov (see Figure \ref{Fig:borisov}) to highlight their relevance to this paper. The vertical dashed line indicates the region interior to which H$_2$O is active, which explains the high inferred C/O ratio of the Centaur 29P. The blue and pink region indicate  primordial cometary C/O ratios indicative of formation interior and exterior to the CO snowline. The comets C/2009 P1 Garradd, C/2010 G2 Hill and C/2006 W3 Christensen  also have high C/O ratios and are indicated with arrows. }
\label{fig:cometsCO}
\end{center}
\end{figure*}

\begin{figure}
\begin{center}
\includegraphics[scale=0.4,angle=0]{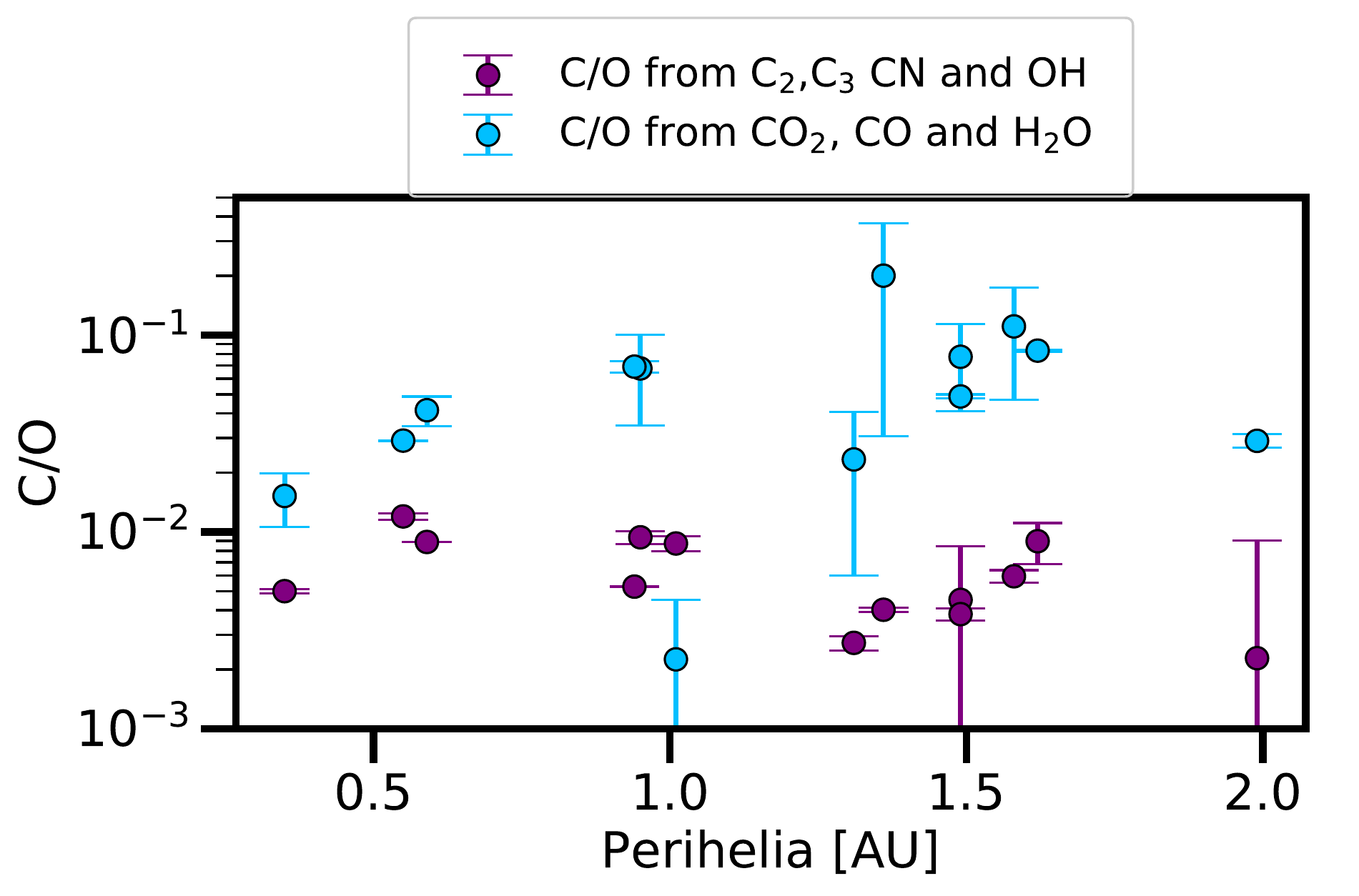}
\caption{A comparison of the C/O ratio of comets inferred from   production rates of CO$_2$, CO and H$_2$O (blue) and C$_2$, C$_3$, CN and H$_2$O from OH (purple). These 13 comets represent the overlap between the comets in  Table \ref{tab:production rates} and those presented in \citet{Ahearn:1995}.   With only one exception, CO$_2$ and CO are the dominant carbon-bearing species for calculating the C/O ratio.}
\label{fig:cometsCO_comparison}
\end{center}
\end{figure}

\begin{figure*}
\begin{center}
\includegraphics[scale=0.55,angle=0]{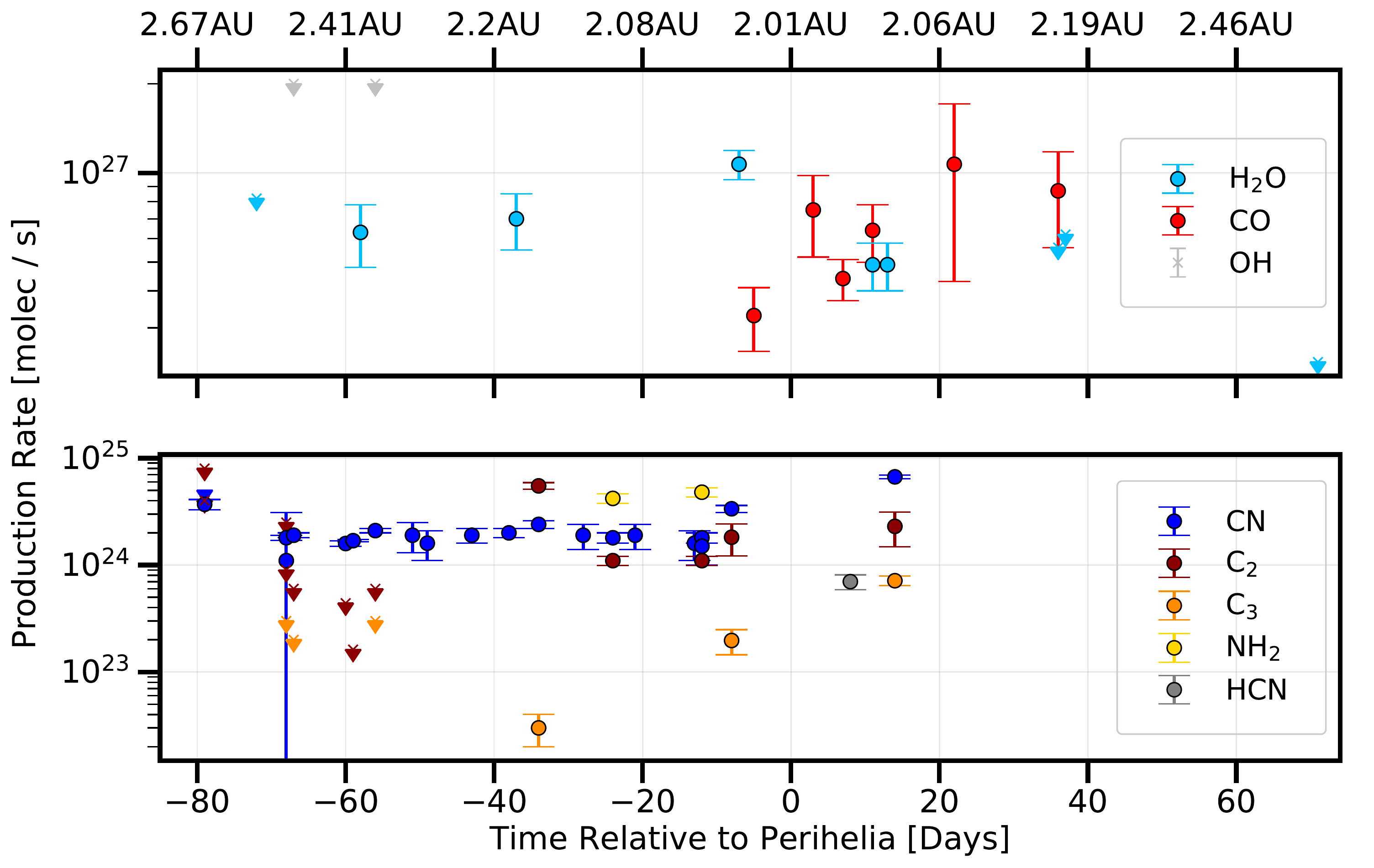}
\caption{The production rate of various volatile species as a function of time relative to perihelia for the interstellar comet 2I/Borisov. The top panel shows the production rates for H$_2$O and CO, which are two orders of magnitude higher than those for CN, C$_2$, C$_3$, NH$_2$ and HCN shown in the bottom panel. We also show upper limits measured for OH in the top panel. These production rates and their respective references are itemized in Table \ref{tab:borisov}. }
\label{Fig:borisov}
\end{center}
\end{figure*}

Toward constraining the C/O in volatiles of known comets, we compiled an extensive list of comets with H$_2$O, CO and/or CO$_2$ production rates measured at some point in their trajectory. In Table \ref{tab:production rates} we show measured  production rate ratios and associated uncertainties of CO$_2$ and CO relative to H$_2$O in this substantially smaller sample of Solar System comets.  We calculate the observed C/O ratio in the coma using the equation, 

\begin{equation}\label{eq:c_to_ratio}
\begin{split}
    {\rm C/O} =\bigg[ \bigg(\frac{{\rm Q(CO)}}{{\rm Q(H_2O)}}\bigg)+ \bigg(\frac{{\rm Q(CO_2)}}{{\rm Q(H_2O)}}\bigg)+ 2\bigg(\frac{{\rm Q(C_2)}}{{\rm Q(H_2O)}}\bigg)\\+ 3\bigg(\frac{{\rm Q(C_3)}}{{\rm Q(H_2O)}}\bigg)+ \bigg(\frac{{\rm Q(CN)}}{{\rm Q(H_2O)}}\bigg)+ \bigg(\frac{{\rm Q(CH_3OH)}}{{\rm Q(H_2O)}}\bigg)\bigg]\bigg/\\\bigg[ 1+ \bigg(\frac{{\rm Q(CO)}}{{\rm Q(H_2O)}}\bigg)+ 2\bigg(\frac{{\rm Q(CO_2)}}{{\rm Q(H_2O)}}\bigg)+ \bigg(\frac{{\rm Q(CH_3OH)}}{{\rm Q(H_2O)}}\bigg)\bigg]\,,
    \end{split}
\end{equation}
where the $1$ in the denominator represents the normalized H$_2$O production rates, specifically ${\rm Q(H_2O)}/{\rm Q(H_2O)}$, and $Q(X)$ is the production rate of a species in units of molecules per second. Note that OH is not included explicitly in Equation \ref{eq:c_to_ratio}, because OH is used to infer the H$_2$O production rate as described previously in this section. Although O$_2$ was detected in 67P  with a mean  abundance  of $3.80 \pm 0.85$ relative to H$_2$O \citep{Bieler2015}, we  neglect its relative contribution to the C/O ratio in Equation \ref{eq:c_to_ratio} because this is not easily detectable remotely.  We calculated the C/O ratio using only the production rates of measured species in Equation \ref{eq:c_to_ratio}.

\begin{figure}
\begin{center}
\includegraphics[scale=0.6,angle=0]{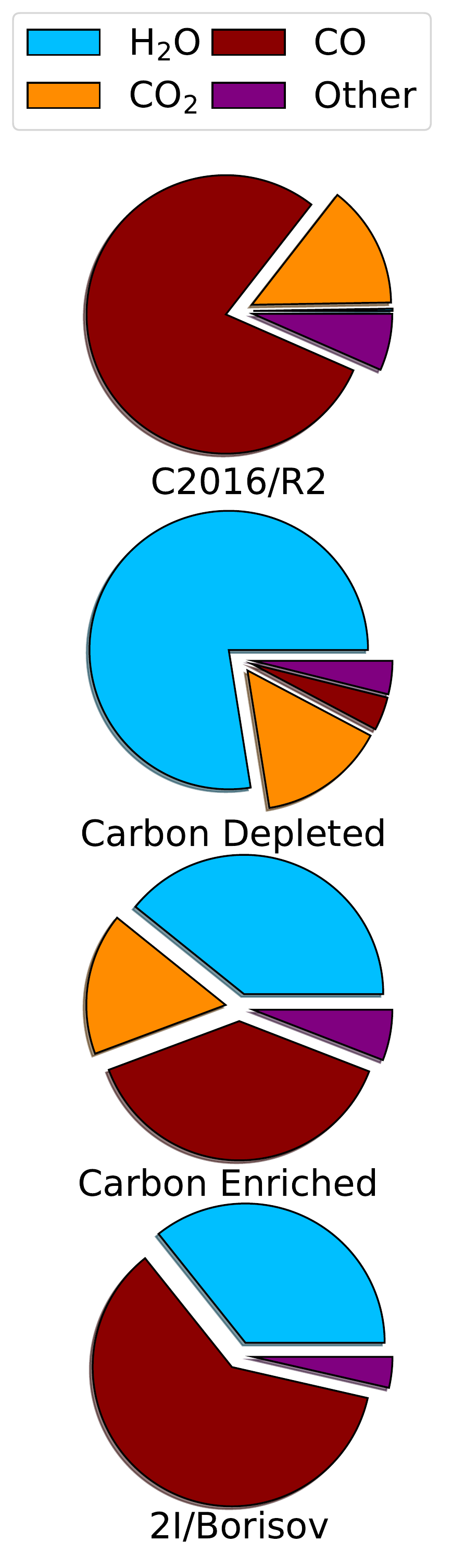}
\caption{A comparison of the composition of R2, 2I, and  typical carbon enriched and  depleted Solar System comets. This figure is a generalized version of the analogous figure in \citet{McKay2019}. It is feasible that R2 and Borisov are representative of comets that formed exterior to the CO snowline in their original proplanetary disks, while  typical Solar System comets formed interior to the CO snowline. The carbon depleted comet is representative of many of the comets in Table \ref{tab:production rates}, and the carbon enriched comet is W3 Christensen. The composition for Borisov is derived from Table \ref{tab:borisov} and the references therein, and  R2 is from \citet{McKay2019}. The lack of CO$_2$ for 2I is only because no measurement of CO$_2$ has been reported. }
\label{fig:R2_2I_pie}
\end{center}
\end{figure}

In Figure \ref{fig:cometsCO}, we show the C/O ratio in the sample of comets presented in Table \ref{tab:production rates}. We also include the sample of comets in Figure \ref{fig:ahearn_comets}, whose C/O is generally much lower because of the lack of measured CO$_2$ and CO production rates. We also include a sample of extrasolar planetesimals from polluted white dwarfs, which reflect the bulk composition of a recently accreted planetesimal. These data include 16 polluted white dwarfs from \citet{Xu2013}, \citet{Farihi2013}, \citet{Xu2014}, \citet{Wilson2015} and \citet{Wilson2016} (see Table 3 in \citet{Wilson2016}).   The polluted white dwarf events all have inferred C/O $<0.5$, and all but two have C/O$<0.1$. 

It is important to note that the C/O derived from volatile production rates  based on coma observations neglect grains that would otherwise contribute to the bulk C/O \citep{Combi2020,Hoang2020_rosetta,Hoang2019_rosetta}. For example, \citet{rubin2019} pointed out the substantial refractory inventory of carbon in dust grains and organics of 67P/Churyumov-Gerasimenko. The oxygen inventory of dust grains also rivals that of ices (Table 4 in \citet{rubin2019}); but both depend on the dust-to-ice weight ratio, which was assumed to be around $1-3$. Here we only focus on the volatile C/O, which is representative the  C/O of gas and ices in the formation location of the protoplanetary disk. 

There are a small subset of objects that have C/O $\ge0.5$. The most notable objects are  2I/Borisov, R2 and `Oumuamua.  We show four post perihelia measurements of Borisov's  C/O ratio  and a nominal C/O value for `Oumuamua, under the assumption that its non-gravitational acceleration was driven by outgassing of CO \citep{Seligman2021}. The Centaur 29P also has a high C/O $>0.5$, but its large heliocentric distance $\gtrsim 6$~au implies that there is  likely more H$_2$O present in the comet that is inactive.  R2 and 2I are atypical with respect to the bulk composition of  every other cometary object that has had their production rates measured.  It is worth noting that the comets C/2009 P1 Garradd, C/2010 G2 Hill and C/2006 W3 Christensen (labeled in Figure \ref{fig:cometsCO})  also have high C/O ratios, which may also imply distant formation location.  However, these C/O ratios are still lower than those of R2 and 2I. 

In Figure \ref{fig:cometsCO_comparison}, we show estimated C/O ratios for  comets that  have measured production rates  for H$_2$O, CO, CO$_2$, C$_2$, C$_3$ and CN. Consistently, with only one exception, the inferred C/O is larger by about an order of magnitude when calculated from CO$_2$ and CO production rates compared to when calculated from C$_2$, C$_3$ and CN. 
This implies that observations of interstellar comet production rates should prioritize H$_2$O, CO$_2$ and CO, in order to estimate the volatile C/O ratio as a tracer for formation location. However, measurements of any carbon or oxygen bearing species would  improve the estimates of the C/O ratio.  The comets in Figure  \ref{fig:cometsCO_comparison} are 1P/Halley 1682 Q1, 8P/Tuttle 1858 A1, 19P/Borrelly 1904 Y2, 103P/Hartley 2 1986 E2, 67P/Churyumov-Gerasimenko 1969 R1, 22P/Kopff 1906 Q1, 81P/Wild 2 1978 A2, 88P/Howell 1981 Q1, 9P/Tempel 1 1867 G1, 116P/Wild 4 1990 B1, C/Bradfield 1979 Y1, C/Austin 1989 X1 and C/Levy 1990 K1.

\subsection{The CO-enriched Comets}

In Table \ref{tab:borisov} we provide a review of all of the production rates for various species that were measured for 2I/Borisov.  In Figure \ref{Fig:borisov}, we show all of these production rates, as function of date and heliocentric distance. While CN, C$_2$, C$_3$, NH$_2$ and HCN were detected in the coma, the activity was dominated by H$_2$O and CO. Observations sensitive to CO were only obtained after perihelion, so it is feasible that there was significant production of CO prior to perihelia.  

The compositions of R2 and 2I are indicative of formation exterior to the CO snow line. In Figure \ref{fig:R2_2I_pie}, we show comparisons of the bulk composition of typical carbon enriched and carbon depleted Solar System comets, R2, and 2I. The differences between these four compositions are striking, especially when viewed in a pie chart. The orders of magnitude higher abundance of CO ice than H$_2$O in R2 is difficult to explain with typical cometary formation scenarios \citep{Mousis2021}. 

In any case, it is feasible  that  planetary systems  typically produce two distinct populations of comets: CO-enriched objects  exterior to the CO snowline and CO-depleted objects  interior to the CO snowline. The existence of a single interstellar comet with a high measured C/O ratio may imply that comets that form exterior to the CO snowline are more readily ejected, while CO-depleted objects are more likely to remain gravitationally bound to the star. The validity of this hypothesis  will be revealed when more interstellar comets are detected and the fraction of the population that are CO-enriched is better constrained.

It is worth noting that cometary production rates and ratios of different species  change as a function of age and position in their trajectory. Given this, we advocate for multiple observations of the relevant production rates at various heliocentric distances for future interstellar comets. We discuss this further in \S \ref{sec:processing} subsection 5.4 and \S \ref{sec:observations}.

\subsection{Implications for Small Body Formation Efficiency}

The host system of an interstellar comet cannot  be determined from its trajectory. As we show in this subsection, however,  the composition of an interstellar comet provides some constraints on the host system. We calculate the total reservoir of disk material interior and exterior to the CO ice line as a function of  stellar mass in order to provide some constraints on the host system. From these calculations, we provide constraints on the types of stars that can produce CO-enriched comets like Borisov. 

 We constructed a grid of circumstellar disk models and calculated the fraction of total material  predicted to be CO-enriched. We set the inner and outer disk boundaries for each disk at 10 times the stellar radius and 100$\,\text{au}(M/M_{\odot})^{1/3}$, respectively. In addition, we set the CO ice line at $30\,\text{au}(M/M_{\odot})^{l/2}$, where $l \sim 3.5$ is the exponent in the mass-luminosity relationship $L\propto M^{l}$. The surface mass density is approximated with a power-law $\Sigma = \Sigma_{o} r^{-\alpha}$. Although the classic minimum mass solar nebula construction adopted $\alpha = 1.5$ \citep{weidenschilling1977mmsn,hayashi1981nebula}, we allow for $\alpha$ to vary.

\begin{figure}
\begin{center}
\includegraphics[scale=0.45,angle=0]{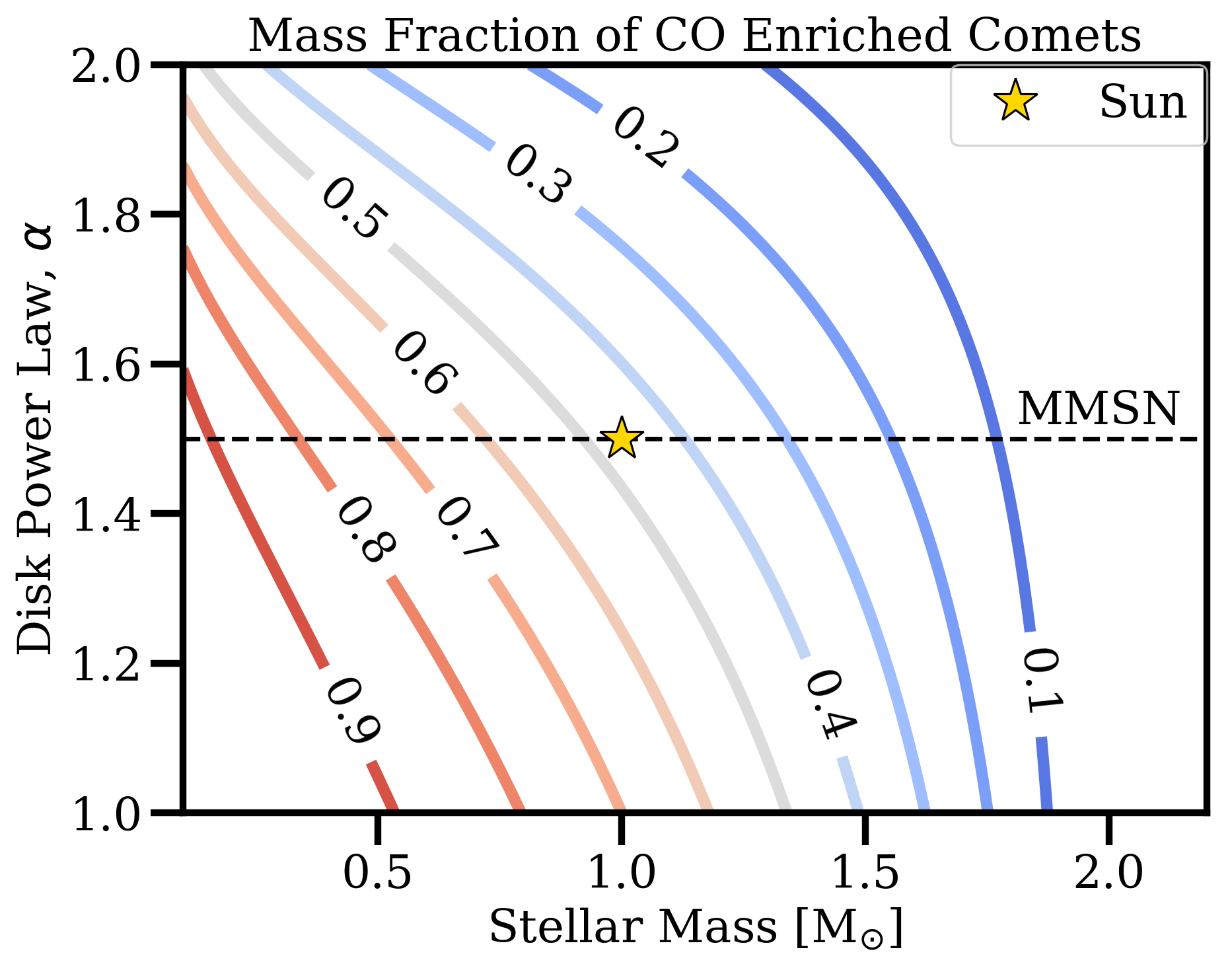}
\caption{The fraction of mass located exterior to the CO ice line in circumstellar disks for a range of stellar masses and assigned surface density profiles. The dotted line indicates the classically adopted power-law exponent for the minimum mass solar nebula.}\label{Fig:nebula}
\end{center}
\end{figure}

From these values, we  compute the fraction of material in the circumstellar disk that is CO-enriched and CO depleted. We assume that  the CO-enriched fraction is the percentage of mass that is located exterior to the CO ice line. Figure \ref{Fig:nebula} shows the mass fraction of each disk that is CO-enriched. More massive stars push $d_{\text{CO}}$ to farther radii and have a lower fraction of CO-enriched material. In addition, larger $\alpha$ leads to a steeper decline in the disk surface density and also decreases the fraction of material exterior to the CO ice line.

As a point of reference, a Solar-mass star would have 70\%, 46\%, and 15\% of its disk mass exterior to the CO ice line for $\alpha = $ 1.0, 1.5, and 2.0, respectively. A 0.08M$_{\odot}$ M-dwarf and a 1.5M$_{\odot}$ A-dwarf would have approximately 91\% and 28\% of their disk masses exterior to their CO ice lines for $\alpha = 1.5$.

While this calculation provides an estimate of the fraction of CO-enriched material as a function of stellar and  disk properties, the  ejection efficiencies of CO-enriched comets  will depend on the cometary formation efficiency and the ejection mechanisms and rates.  The propensity for stars of various masses to eject CO-enriched comets will depend on the typical outcomes of planet formation as a function of stellar mass. One possibility is that planets with large Safronov number (the ratio of the escape velocity to the orbital velocity) form at large semimajor axes for M-dwarf stars, although this idea has been theoretically disfavored \citep{laughlin2004giantplanet}. \textit{If the CO-enriched composition of Borisov is representative of typical compositions of interstellar comets, then it seems unlikely that stars larger than the Sun contribute substantially to the overall population. }

\section{The Relative Importance of Erosion in the Solar System and the Interstellar medium}\label{sec:processing}
A confounding mechanism that must be taken into account when predicting the molecular composition of an interstellar comet is that these objects  experience  processing in the interstellar medium. The comet then experiences additional processing 
from stellar irradiation in the Solar System. The relative importance of these two processes effects our ability to measure the \textit{primordial}\footnote{Primordial is loosely defined as representative of the composition upon ejection from the host system, although there is likely additional processing prior to ejection. In this paper, we do not attempt to model the pre-ejection processing due to the uncertainty in host system properties and a comet's lifetime before ejection.} composition. If the object experiences drastically more processing in the ISM than in the Solar System, then  measured production rates will not be representative of the primordial composition. On the other hand, if an object experiences more processing in the Solar System than in the ISM, then the measured production rates (especially post perihelia) will more directly probe the primordial composition.  In this section, we quantify the relative importance of these two effects for a range of interstellar comet lifetimes and trajectories. 

An interstellar comet is exposed to isotropic  radiation  in the interstellar medium, primarily from  cosmic rays but also from intermittent background stellar radiation. Surface volatile material will undergo non-thermal cosmic ray-induced desorption and desorption via the absorption of FUV photons \citep{Hollenbach_2008}, which  causes continuous erosion. Ice mantled grains are transiently heated from cosmic rays and cool to their equilibrium temperature via sublimation of  ice. This  process was considered in Section 3 of \citet{seligman2020}, \citet{Levine2021_h2} and \citet{hoang2020destruction} for an H$_2$ enriched  `Oumuamua and  is similar to erosion via particle bombardment  \citep{Domokos2009,Domokos2017}. The processing of interstellar objects composed of N$_2$, CO, CO$_2$, and CH$_4$ due to both cosmic ray-induced desorption and collisional heating with ambient gas in the interstellar medium was investigated by \citet{Phan2021}. They estimated that a km-scale progenitor could survive in the galactic cosmic ray environment for $\sim1\times10^9$, $\sim1\times10^9$, $\sim2\times10^9$, $\sim5\times10^9$ years if it was composed of N$_2$, CO, CH$_4$ and CO$_2$, respectively. These values represent upper limits, because the cosmic rays will dissociate molecules in addition to causing ices to desorb. 

It is worth noting that the analysis that follows also applies to Oort Cloud comets. The Oort Cloud is located  beyond the heliopause, with  environmental conditions similar to the interstellar medium. 

\subsection{Numerical Simulations}

In this subsection, we perform numerical simulations tracking the aspect ratio, size, and density of a km-scale interstellar comet like 2I/Borisov during the past 10 GYRs of evolution. We  backtrace the trajectory from initial assumptions about the composition, bulk density and size when the object enters the Solar System, and we calculate the geometric changes from continuous desorption from cosmic rays and intermittent FUV photons\footnote{An augmentation of the value of  $\Phi_{CR}$ could be applied to incorporate the  effect of intermittent  desorption via the absorption of FUV photons, but  the qualitative effects of modelling both of these processes  together are similar to modelling only one.}. The nominal Solar System entrance is defined as the heliocentric distance when the change in radius becomes asymptotic.

We adopt three initial compositions for a comet  composed entirely of (i) H$_2$O, (ii) CO$_2$, and (iii) CO ice with a roughly spherical nucleus with a $1$km diameter and aspect ratio of $1:1:0.98$. The bulk density of the comet is assumed to be the same as that of  the ice, although it is important to note that solar system comets are porous with typical bulk densities $<0.5$g/cm$^3$. However, the \textit{relative} mass loss of different volatiles does not sensitively depend on the assumed bulk density. 

At each timestep of length $\Delta \tau$ in the simulation, we calculate the addition of an ellipsoidal shell of material desorbed from the surface. The time when the comet enters the Solar System is  $t_0$ , and positive/negative $\Delta \tau$ corresponds to time intervals moving forward/backward in time for  the remainder of these calculations. The surface integrated production rate, ${\cal N}$, of sublimated molecules for a given species, $X$ per unit time, is given by the equation,
\begin{equation}\label{eq:prodrate}
{\cal N}(t,X)=+\frac{\Sigma(t) \Phi_{CR}}{\Delta H(X)/N_{A}+\gamma k\textit{T}_{\rm Sub}(X)}\, ,
\end{equation}
where $\Delta H(X)$ is the molar enthalpy of sublimation  of the volatile species $X$, $\textit{T}_{\rm Sub}(X)$ is its sublimation temperature, $\Sigma$ is the ellipsoidal surface area at time $t$, and $\gamma$ is the adiabatic index of the escaping vapor.  This calculation assumes that $100\%$ of the energy received is deposited into desorption of ice via sublimation and neglects the  contribution to radiative cooling of the grain via blackbody emission (which is negligible at ISM temperatures).  

The change in the comet's mass, $\delta m$, during each time step of length $\Delta \tau$ is given by multiplying Equation \ref{eq:prodrate} by the mass of the species and the time step length, $\delta m=-\mu m_{\rm u}{\cal N}\Delta \tau$, for a species with mass $\mu m_{\rm u}$, where $ m_{\rm u}$ is the atomic mass constant. The negative sign ensures that mass is added/removed from the body as the integration moves backwards/forwards in time. The resulting change in volume, $\delta V$, is given by  $\delta V=\delta m/\rho$, where $\rho$ is the bulk density of the volatile species that is sublimated.  In order to  explicitly demand mass conservation, we solve for roots of the cubic equation, $\zeta(\delta h)$, where $\delta h$ is the change in length of each principal axis denoted $a,b$ and $c$,
\begin{equation}
     \zeta(\delta h)=\delta V-\frac43\pi\bigg( (a+\delta h)(b+\delta h)(c+\delta h)-abc\bigg)\,.
\end{equation}
% \begin{equation}
%      \zeta(\delta h,\mp\Delta\tau)=\delta V-\frac43\pi\bigg( (a\pm\delta h)(b\pm\delta h)(c\pm\delta h)-abc\bigg)\,.
% \end{equation}
The signs of $\delta V$ and $\delta h$ depend on whether the integration is backwards ($\delta V>0$, $\delta h>0$) or forwards ($\delta V<0$, $\delta h<0$) in time. We verified that the results were not sensitive to the initial aspect ratios. We show the  thermodynamic properties for the relevant species in Table \ref{table:thermo}. 

\begin{figure}
\begin{center}
\includegraphics[scale=0.4,angle=0]{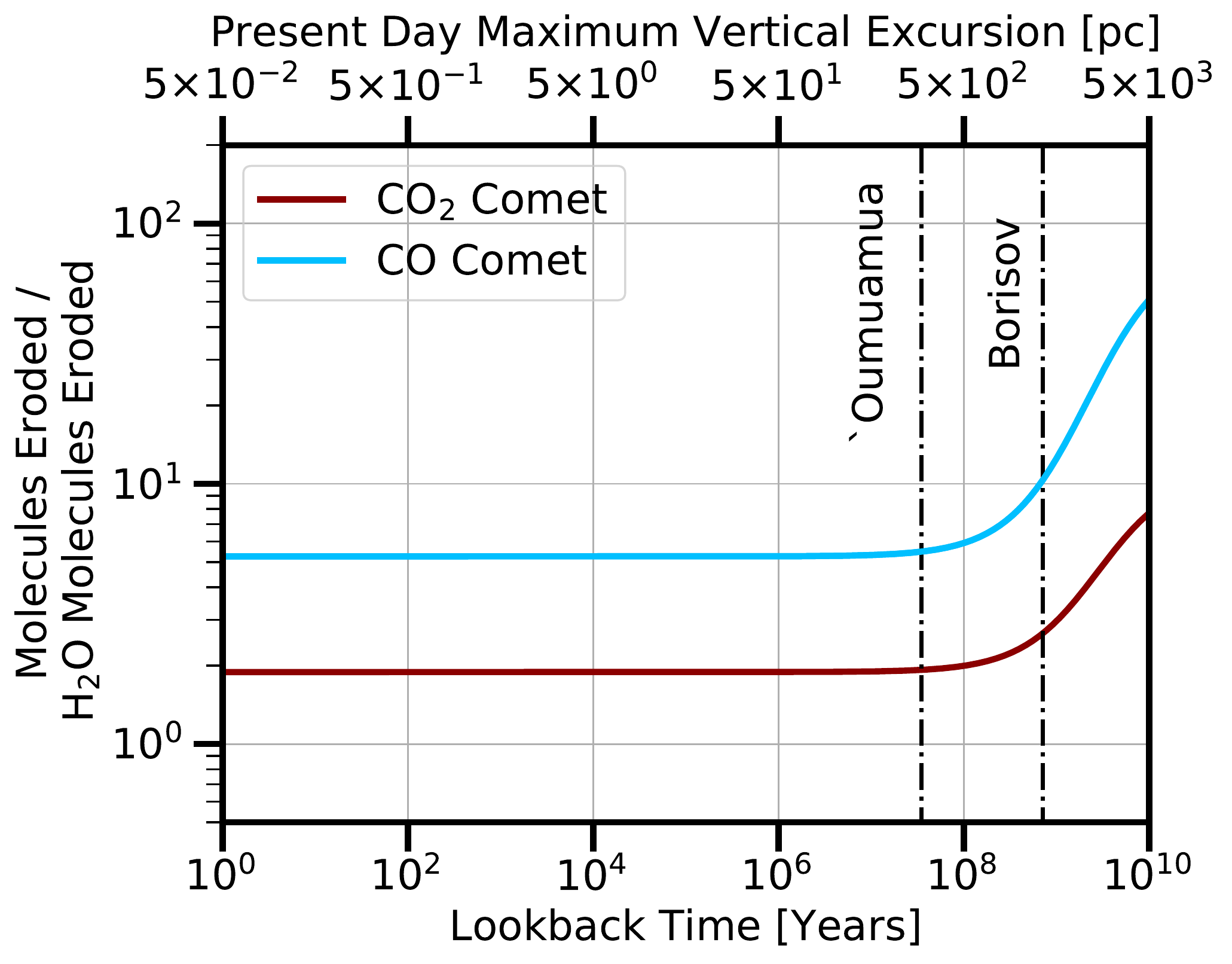}
\caption{The evolution of relative production rates of a km-scale comet experiencing mass loss in the interstellar medium. The blue and red curves indicate  the  time-integrated number of sublimated CO and CO$_2$ relative to H$_2$O molecules.  The lower x- axis corresponds to the lookback time \textit{prior} to entering the Solar System, while the upper axis shows the corresponding \textit{present day} maximum vertical excursion in the galactic orbit. The dashed-dotted lines indicate the inferred age and excursion of `Oumuamua and Borisov. }\label{Fig:productionrate}
\end{center}
\end{figure}

The relative mass loss of CO and CO$_2$ with respect to H$_2$O are shown in Figure \ref{Fig:productionrate}.
The top axis indicates the \textit{present day} maximum vertical excursion in the galactic orbit corresponding to the age on the lower x-axis. This approximates the vertical velocity dispersion, and therefore age, due to dynamical heating and primordial dispersion. `Oumuamua and Borisov  had ages close to $\tau_{1I}\sim35$ Myr and $\tau_{2I}\sim710$ Myr, given that the maximum  vertical excursion $z_{ISO}\sim\sqrt{\tau}$ \citep{Hsieh2021}. The  H$_2$O/CO$_2$/CO comet grew to  diameters of $\sim1.004/1.010/1.021$km, $\sim1.400/1.963/3.078$km and $\sim4.958/10.629/21.781$km after $10^7$, $10^9$ and $10^{10}$ years of (backwards) evolution. The evolution is not constant after $>10^8$ years, when the larger  surface areas of the CO$_2$ and CO comet relative to the H$_2$O comet significantly enhance the intercepted energy flux, so these ratios should be interpreted as  upper limits. %In the following subsection, we introduce an analytic approximation that can be used to infer the geometric changes assuming a mixed composition. 
\setcounter{table}{2}
\begin{table}
\centering
\caption{ The enthalpy of sublimation ($\Delta$ H),  solid density ($\rho$), and temperature of sublimation ($\textit{T}_{\rm Sub}$) of different volatile species. }\label{table:thermo}
\medskip
\begin{tabular}{ccccp{13mm}p{13mm}p{13mm}p{13mm}p{13mm}}
%\begin{tabular}{cccccc}
\hline
Species & $\textit{T}_{\rm Sub}$\, [$\rm{K}$] & $\rho$ \, [${\rm g\, cm^{-3}}$]& $\Delta$ H [${\rm kJ\, mol^{-1}}$]  \\
\hline
 %&  &  &  &  \xi\sim0.20& \xi\sim0.13& \xi\sim0.25& \xi\sim0.25&\\
 CO & 60. & 1.60& 8.1 \\
CO$_2$ & 82. & 1.56& 28.84 \\
H$_2$O & 155. & 0.82& 54.46\\
\hline
\end{tabular}
\end{table}
\begin{figure*}
\begin{center}
\includegraphics[scale=0.2,angle=0]{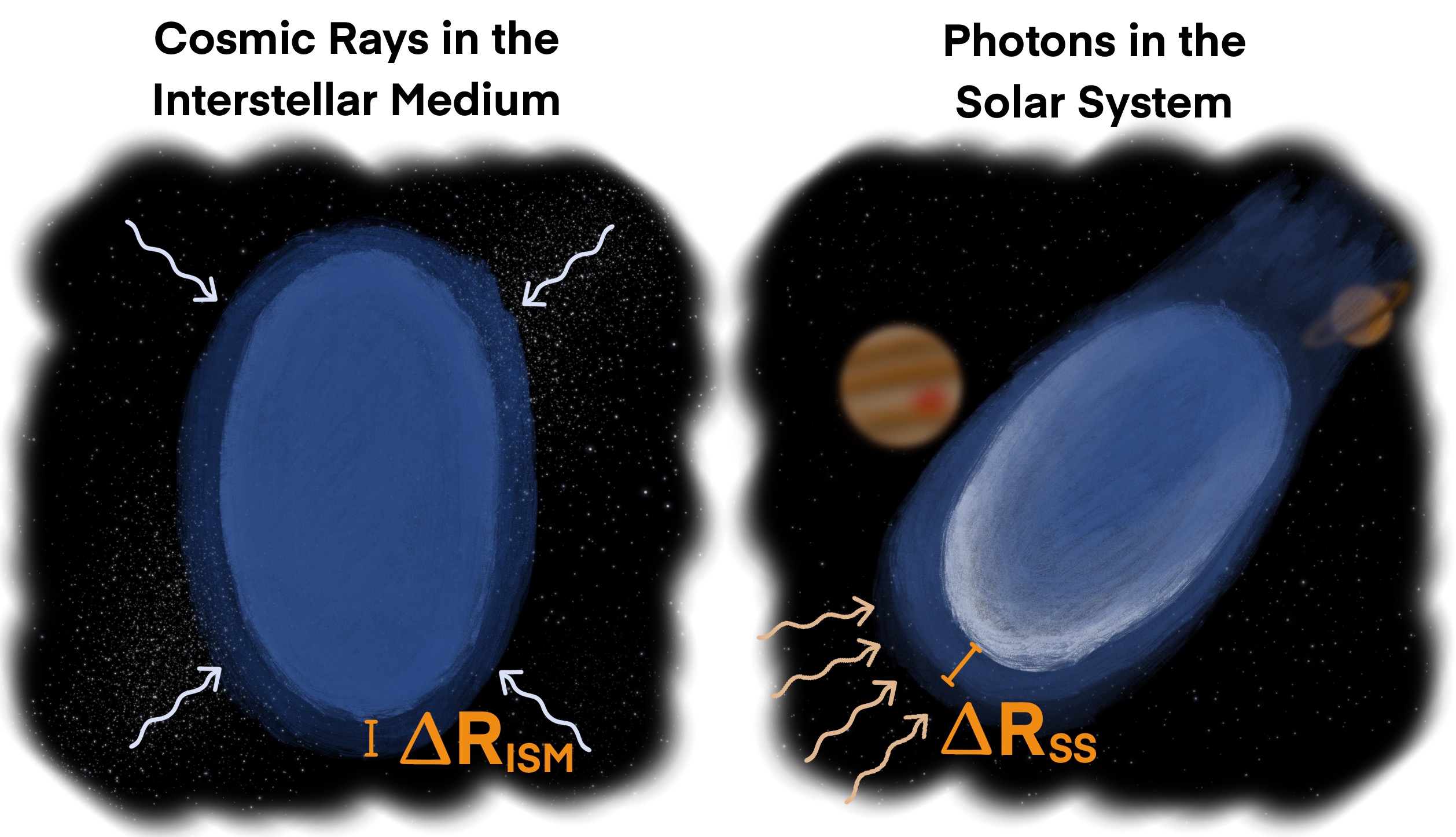}
\caption{A schematic diagram showing two distinct stages of an interstellar comets' lifetime. On the left hand side, the  comet is travelling through the interstellar medium. It experiences  continuous isotropic radiation and interaction with ambient gas, which removes volatile material off of the surface. The overall reduction of the radius in this phase of the object's lifetime is denoted as $\Delta {\rm R}_{\rm ISM}$. On the right hand side, the comet has finished its journey through the interstellar medium and is travelling through the Solar System. There is a focused ablation of material due to sublimation from solar irradiation, which produces a cometary tail. Averaged over many tumbles of the object and orientations to the Sun, this erosion produces an additional change in radius, which we denote as  $\Delta {\rm R}_{\rm SS}$.}\label{Fig:schematic}
\end{center}
\end{figure*}
\subsection{Analytic Approximation for ISM Processing }

In this subsection,  assuming a spherical geometry throughout the evolution, we   derive a simplified analytic  form for the  evolution in the interstellar medium as a function of age. At any time step in the integration, the rate of change of the comet's volume due to the ongoing ablation is given by %the differential volume ablated, $\delta V$ is given by

% \begin{equation}\label{eq:volume_diff}
% \begin{split}
% \delta V(t+\Delta \tau,X)=-3\,\bigg(\,\frac{4\pi}{3}\,\bigg)^{1/3}\bigg(\frac{\mu m_{\rm u}}{\rho}\bigg)\,\\\bigg(\frac{ \Phi_{CR}}{\Delta H(X)/N_{A}+\gamma k\textit{T}_{\rm Sub}(X)}\bigg)\,V(t)^{2/3}\Delta \tau\, ,
% \end{split}
% \end{equation}
\begin{equation}\label{eq:volume_diff}
\begin{split}
\frac{d V(t,X)}{dt}=-3\,\bigg(\,\frac{4\pi}{3}\,\bigg)^{1/3}\bigg(\frac{\mu m_{\rm u}}{\rho}\bigg)\,\\\bigg(\frac{ \Phi_{CR}}{\Delta H(X)/N_{A}+\gamma k\textit{T}_{\rm Sub}(X)}\bigg)\,V(t)^{2/3}\, ,
\end{split}
\end{equation}
where $V(t)$ is the volume of the comet at the time $t$ in the integration. The assumption of spherical symmetry allows for the substitution for surface area $\Sigma = 3V/R = 3(4\pi/3)^{1/3} V^{2/3}$, where $R$ is the radius of the comet.  Equation \ref{eq:volume_diff} can be integrated via the expression,

\begin{equation}\label{eq:integral}
\begin{split}
\int_{V_0}^{V_i}V^{-2/3}\,dV=\\-\int_{t_0}^{t_0-\tau_{ISO}}3\bigg(\frac{4\pi}{3}\bigg)^{1/3}\bigg(\frac{\mu m_{\rm u}}{\rho}\bigg)\bigg(\frac{ \Phi_{CR}}{\Delta H/N_{A}+\gamma k\textit{T}_{\rm Sub}}\bigg) \,dt\, ,
\end{split}
\end{equation}
where $V_0$ represents the volume when it enters the Solar System, while $V_i$ is the volume at the time of ejection. $\tau_{ISO}$ is the duration of the journey in the ISM or ISO age, and is positive.  Equation \ref{eq:integral} can be integrated to give the linear function for the radius,
\begin{equation}\label{eq:diameter}
    R_i=R_0+\,\bigg(\frac{\mu m_{\rm u}}{\rho}\bigg)\,\bigg(\,\frac{ \Phi_{CR}}{\Delta H/N_{A}+\gamma k\textit{T}_{\rm Sub}} \,\bigg)\,\tau_{ISO}\,
\end{equation}
where $R_0$ and $R_i$  are the radius when it enters the Solar System and upon ejection, respectively.

We verified that Equation \ref{eq:diameter} produced good agreement with the numerical calculations shown in Figure \ref{Fig:productionrate} in the previous subsection.  Equation \ref{eq:diameter} can be generalized for any mixture of volatiles as long as the bulk density of the comet remains constant.   After this processing transpires, the outer rind of a comet with a mixture of ices will be devoid of the molecules with lowest $\Delta H$ and $\textit{T}_{\rm Sub}$. The interior of the comet will still retain the composition upon ejection. 

\subsection{Solar System Processing}

In this subsection, we extend the previous calculation to account for processing in the Solar System. Since the interior composition remains intact in the ISM, post perihelia observations will be  representative of the primordial compositions if the Solar System processing removes the ISM-processed surface. 

To quantify this effect, we derive an analytic function for the radius evolution  in the Solar System, analagous to Equation \ref{eq:diameter} for the ISM. A schematic diagram of these two  processes is shown in Figure \ref{Fig:schematic}. For the encounter in the Solar System, the rate of change of the comet's volume is %the differential volume change, $\delta V$ for the mass lost during any $\Delta \tau$ is

% \begin{equation}\label{eq:volume_diff_SS}
% \begin{split}
% \delta V(t+\Delta \tau)=-3\bigg(\frac{4\pi}{3}\bigg)^{1/3}\bigg(\frac{ 1}{(\Delta H/N_{A}+\gamma k\textit{T}_{\rm Sub})}\bigg)\\\bigg(\frac{\mu m_{\rm u}}{\rho}\bigg)\bigg(\frac{L_{\odot}}{4\pi r_H^2}\bigg)\,\xi\,(1-p)\,V(t)^{2/3}\,\Delta \tau\, ,
% \end{split}
% \end{equation}
\begin{equation}\label{eq:volume_diff_SS}
\begin{split}
\frac{d V(t,X)}{dt}=-3\bigg(\frac{4\pi}{3}\bigg)^{1/3}\bigg(\frac{ 1}{(\Delta H/N_{A}+\gamma k\textit{T}_{\rm Sub})}\bigg)\\\bigg(\frac{\mu m_{\rm u}}{\rho}\bigg)\bigg(\frac{L_{\odot}}{4\pi r_H^2}\bigg)\,\xi\,(1-p)\,V(t)^{2/3}\, ,
\end{split}
\end{equation}
where $L_{\odot}$ is the solar luminosity, $\xi$ is the ratio of the projected surface area to the total surface area, p is the bond albedo, and $r_H$ is the heliocentric distance at a given point in the trajectory. $\xi=1/4$ averaged over all projection angles \citep{Meltzer1949}. This value is an upper limit because we do not include radiative cooling, which would decrease the magnitude of the mass lost in the Solar System\footnote{The re-radiation efficiency for `Oumuamua was calculated in Figure 1 of \citet{Seligman2018}, and did reach close to $100\%$ for a short timespan surrounding perihelia.}. 

From the conservation of  angular momentum, we have $r_H^2\dot{\theta}=bv_\infty$, where $\theta$ is the angle from the perihelia in the plane of the hyperbolic trajectory, $v_\infty$ is the hyperbolic excess velocity, and $b$ is the impact parameter.  %Substituting $\Delta \tau/r_H^2=d\theta/(bv_\infty)$ into Equation \ref{eq:volume_diff_SS} yields,
Combining $\dot{\theta}=bv_\infty/r_H^2$ with Equation \ref{eq:volume_diff_SS} to change integration variables from $t$ to $\theta$ leads the $r_H^2$ factors to conveniently cancel and yields,

\begin{equation}\label{eq:integral_SS}
 \begin{split}
\int_{V_0}^{V_f}V^{-2/3}\,dV=-\int_{\theta_{min}}^{\theta_{max}}3\bigg(\frac{4\pi}{3}\bigg)^{1/3}\bigg(\frac{\mu m_{\rm u}}{\rho}\bigg)\,\bigg(\frac{1}{bv_\infty}\bigg)\\\bigg(\frac{ 1}{(\Delta H(X)/N_{A}+\gamma k\textit{T}_{\rm Sub}(X))}\bigg)\,\bigg(\frac{L_{\odot}}{4\pi }\bigg)\,\bigg(\xi(1-p)\bigg)\,d\theta\, ,
\end{split}
\end{equation}
where $V_f$ is the final volume after the Solar System encounter.  The  angle between the asymptotes for a hyperbolic orbit is given by\footnote{The inverse cosine function in Equation~\ref{eq:theta_minmax} is restricted to the principal branch such that $\theta_{max}-\theta_{min}$ ranges between $\pi$ and $2\pi$.}
\begin{equation}
\label{eq:theta_minmax}
    \theta_{max}-\theta_{min}=2\cos^{-1}(-1/e)\, ,
\end{equation} where the eccentricity of the trajectory is in turn related to $b$ and $v_{\infty}$,
 
 \begin{equation}
     e^2=1+\frac{b^2v_{\infty}^4}{G^2M_\odot^2}\,.
\end{equation}
Equation \ref{eq:integral_SS} integrates to,
\begin{equation}\label{eq:diameter_SS}
\begin{split}
    R_f=R_0-2\,\bigg(\frac{\mu m_{\rm u}}{\rho}\bigg)\bigg(\,\frac{ 1}{\Delta H/N_{A}+\gamma k\textit{T}_{\rm Sub}} \,\bigg)\,\\\bigg(\frac{L_{\odot}}{4\pi }\bigg)\,\bigg(\xi(1-p)\bigg)\,\bigg(\frac{1}{bv_\infty}\bigg)\,\bigg(\cos^{-1}\big(-\frac{1}{e}\big)\,\bigg)\,,
\end{split}
\end{equation}
where $R_f$ is the final radius of the object after the encounter with the Solar System.  

\subsection{The Relative Importance of Processing in the Interstellar Medium and the Solar System }

In the previous two subsections, we calculated the change in radius of an interstellar comet due to (i) non-thermal cosmic ray-induced desorption in the interstellar medium (Equation \ref{eq:diameter}) and (ii) stellar irradiation during the encounter with the Solar System (Equation \ref{eq:diameter_SS}). In this subsection, we evaluate the relative importance of these two processes as a function of the age and trajectory of an interstellar comet. We define the change in radius in the ISM and Solar System as $\Delta R_{\rm ISM}=R_i-R_0$ and $\Delta R_{\rm SS}=R_0-R_f$. The relative change in radius  is given by the dimensionless quantity, 

% \begin{equation}\label{eq:dd_nonumbers}
% \begin{split}
%   \frac{\Delta {R}_{\rm SS}}{\Delta {R}_{\rm ISM}}=\frac{1}{2\pi}\bigg(\frac{L_{\odot}}{\Phi_{CR}}\bigg) \,\bigg(\xi(1-p)\bigg)\,\\\bigg(\frac{1}{bv_\infty}\bigg)\,\bigg(\cos^{-1}\big(-\frac{1}{e}\big)\,\bigg)\,\bigg(\frac{1}{\tau_{ISO}}\bigg)\,.
%   \end{split}
% \end{equation}

\begin{equation}\label{eq:dd_nonumbers}
  \frac{\Delta {R}_{\rm SS}}{\Delta {R}_{\rm ISM}}=\frac{\cos^{-1}\big(-1/e\big)}{2\pi}\bigg(\frac{L_{\odot}\,\xi(1-p)}{\Phi_{CR}\,b\,v_\infty \tau_{ISO}}\bigg) .
\end{equation}
Compositional measurements will be representative of the primordial material for objects where  $ \Delta {R}_{SS}/\Delta {R}_{ISM}\ge1$.  Equation \ref{eq:dd_nonumbers} can be written as the following scaled relationship,
\begin{equation}\label{eq:dd}
\begin{split}
  \frac{\Delta {R}_{\rm SS}}{\Delta {R}_{\rm ISM}}=
  0.59\bigg(\frac{1-p}{0.9}\bigg)\bigg(\frac{0.85 {\rm au}}{b}\bigg)\bigg(\frac{26{\rm km/s}}{v_\infty}\bigg)\\\bigg(\frac{\cos^{-1}(-1/e)}{\cos^{-1}(-1/1.2)}\bigg)\bigg(\frac{3.5\times10^8{\rm yr}}{\tau_{ISO}}\bigg)\,,
\end{split}
\end{equation}
where values of $b$, $v_\infty$, $e$ and $\tau_{ISO}$ of `Oumuamua have been used.
% This dimensionless number  is independent of the composition.  The eccentricity of the trajectory is formally given by $b$ and $v_{\infty}$,
 
%  \begin{equation}
%      e^2=1+\frac{b^2v_{\infty}^4}{G^2M_\odot^2}\,.
%  \end{equation}

\begin{figure}
\begin{center}
\includegraphics[scale=0.45,angle=0]{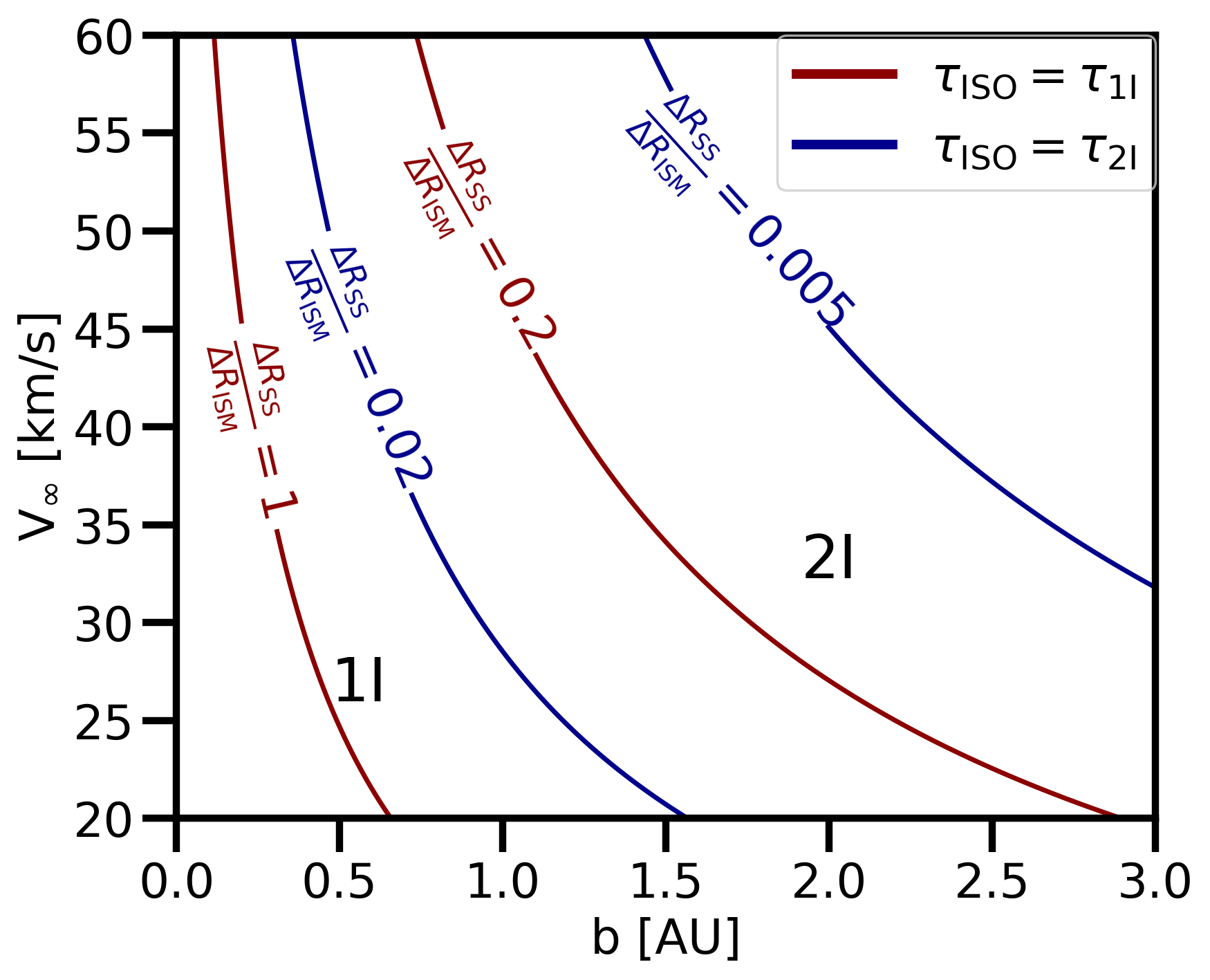}
\caption{The relative erosion of an interstellar comet in the Solar System and the interstellar medium. We show the change in radius of the comet in the Solar System divided by the change in radius in the ISM, calculated for a range of impact parameters and hyperbolic velocities  using Equation \ref{eq:dd}. The ratio for objects the age of 1I/`Oumuamua and 2I/Borisov are shown in red and blue contours. The locations of these two detected interstellar objects are indicated.}\label{Fig:deltaD}
\end{center}
\end{figure}

\begin{figure}
\begin{center}
\includegraphics[scale=0.4,angle=0]{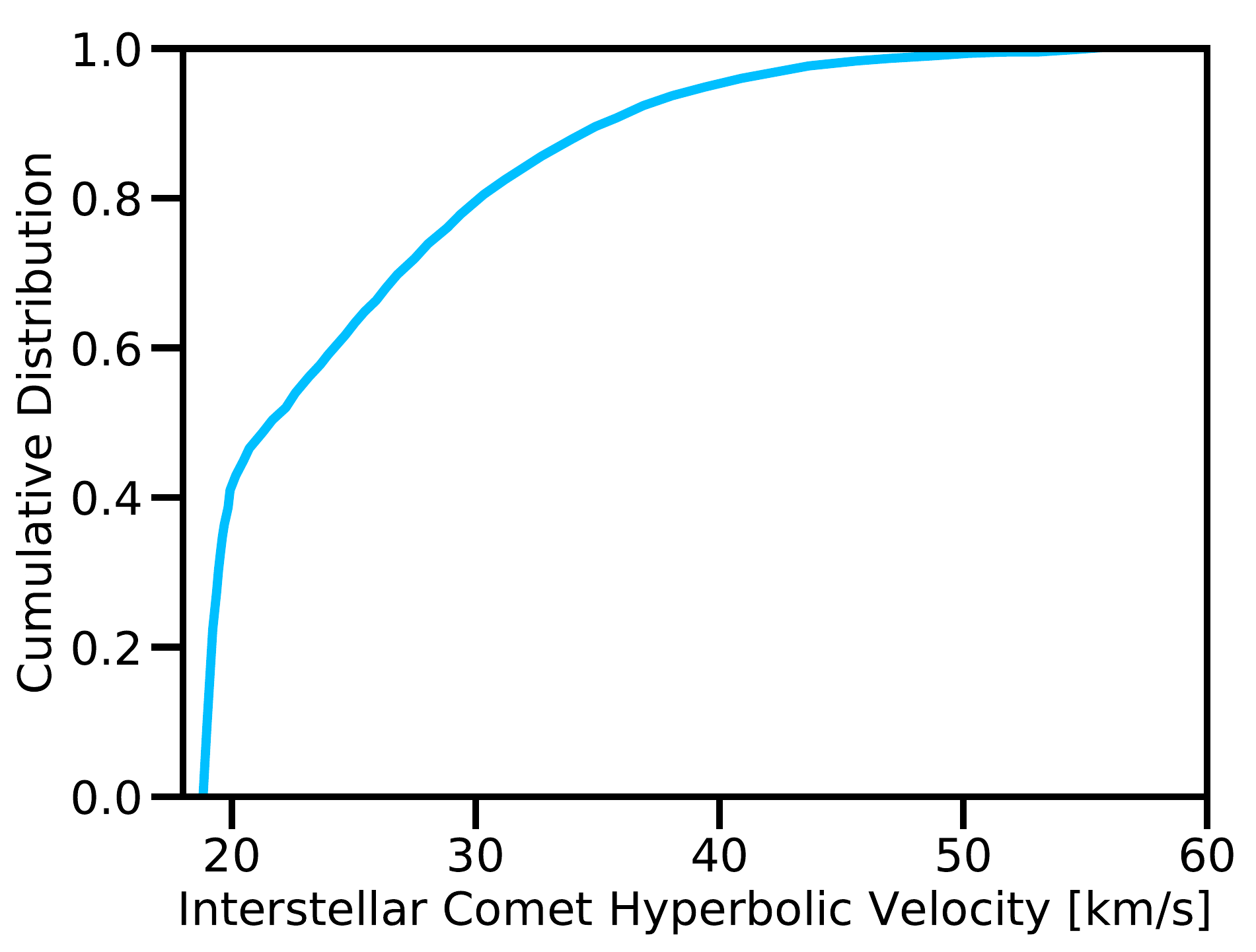}
\includegraphics[scale=0.4,angle=0]{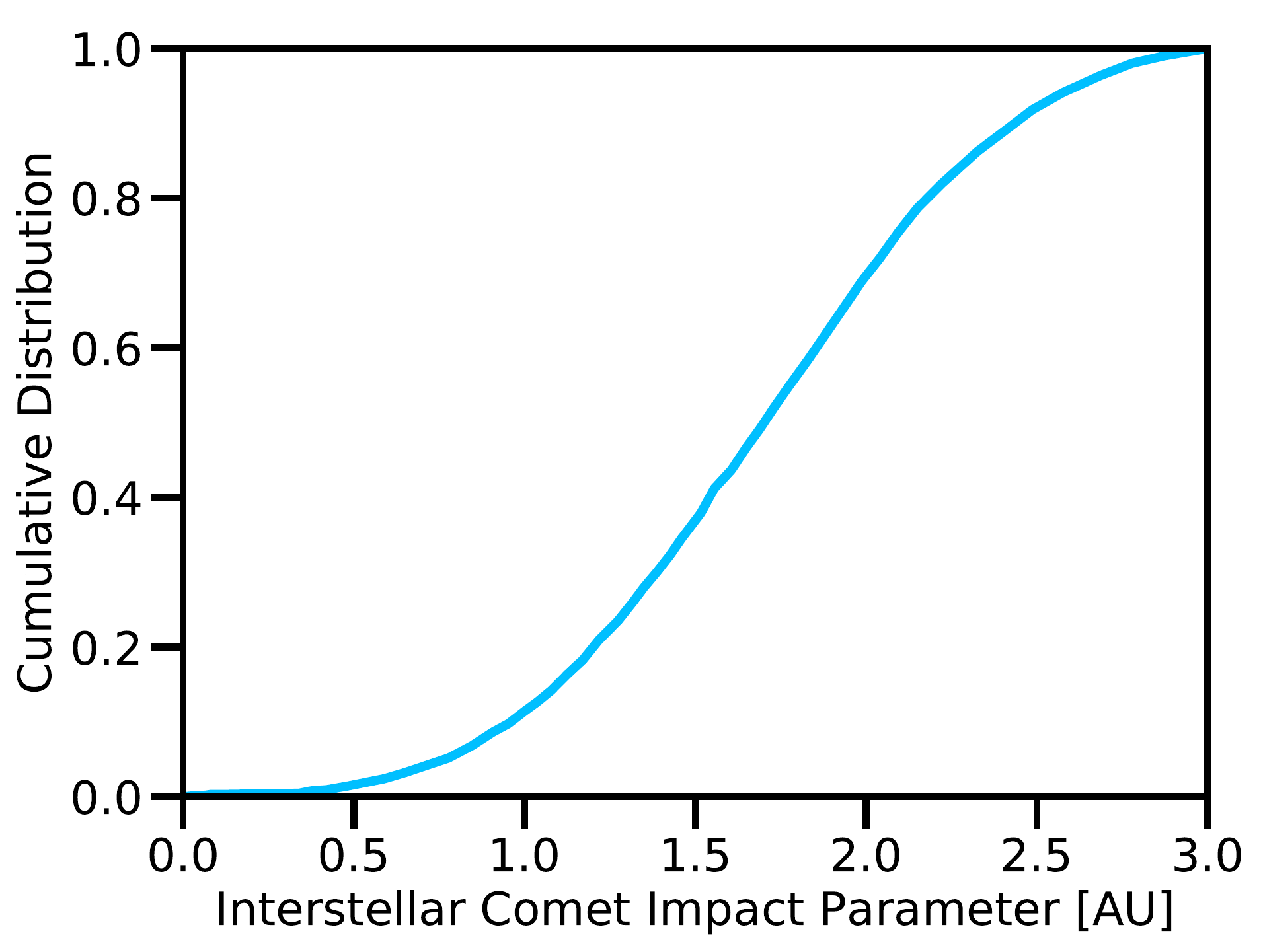}
\caption{The cumulative distribution functions for impact parameter and hyperbolic excess velocity for interstellar objects that will be detected by the LSST, from Figures 12 and 15 in \citet{Hoover2021}. These are calculated using a detailed population synthesis  assuming interstellar objects were as bright as 'Oumuamua. }\label{Fig:cdf_v_b}
\end{center}
\end{figure}

 We show the relative change in radius in the Solar System and ISM  for a range of $b$ and $v_{\infty}$ in Figure \ref{Fig:deltaD} for  Borisov and `Oumuamua aged objects. `Oumuamua experienced comparable erosion in the Solar System and the ISM with $\Delta {R}_{\rm SS}/\Delta {R}_{\rm ISM}=0.59$, while Borisov had  $\Delta {R}_{\rm SS}/\Delta {R}_{\rm ISM}<10^{-2}$. Previously, \citet{Kim2020} estimated that Borisov only lost $\sim0.4$m of surface material in the Solar System and similarly concluded that compositional observations were not representative of the primordial material. \citet{Hoover2021} presented the orbital distribution of interstellar objects detected by LSST, assuming that all of the objects had absolute magnitude similar to that of 'Oumuamua. We show the cumulative distribution functions for  $b$ and $v_{\infty}$ in Figure \ref{Fig:cdf_v_b}, from Figures 12 and 15 of \citet{Hoover2021}. LSST will detect objects out to $b\sim3$au  with $20$ km/s$<v_\infty<60$km/s. These  limits are reflected in Figure \ref{Fig:deltaD}. Roughly $\sim80\%$ of detected objects will have $v_\infty<40$km/s, and $b<2$au.  Objects the age of `Oumuamua  will  have  $\Delta {R}_{\rm SS}/\Delta {R}_{\rm ISM}\sim1$ only for $b\le0.5$au. No objects the age of Borisov will have $\Delta {R}_{\rm SS}/\Delta {R}_{\rm ISM}\ge1$. From incorporating the results of the marginal cumulative distribution functions, we calculate that   $<5\%$ of objects with ages similar to `Oumuamua will have $\Delta {R}_{\rm SS}/\Delta {R}_{\rm ISM}>1$. Spectroscopic measurements during a  tidal disruption event,  activity driven disintegration event, or a space based impactor collision  could reveal the primordial composition. %Therefore,   spectroscopic abundance measurements of interstellar comets should ideally be  obtained  for each object at multiple heliocentric distances.

\subsection{Estimating Primordial Composition From Observed Production Rates}
In this subsection, we outline a method to approximate the primordial C/O ratio from the observed production rates. As we showed in the previous subsection,  the majority of interstellar comets will not exhibit activity representative of their primordial composition (unless they break apart under the action of tides, activity, rotation, and/or collisions). Moreover, processing in the interstellar medium should preferentially remove CO and CO$_2$ with respect to H$_2$O (Figure \ref{Fig:productionrate}). Therefore, the C/O ratio inferred from H$_2$O, CO$_2$ and CO production rates after processing are  lower limits on the  primordial C/O ratio. 

 The volatile C/O ratio of an interstellar comet that has production rates  for CO$_2$, CO and H$_2$O measured during its apparition is given by

\begin{equation}\label{eq:ctoO}
\begin{split}
    \bigg({\rm C/O}\bigg)_{\rm Obs} =\bigg[ \bigg(\frac{{\rm Q(CO)}}{{\rm Q(H_2O)}}\bigg)_{\rm Obs}+ \bigg(\frac{{\rm Q(CO_2)}}{{\rm Q(H_2O)}}\bigg)_{\rm Obs}\bigg]\bigg/\\\bigg[ 1+ \bigg(\frac{{\rm Q(CO)}}{{\rm Q(H_2O)}}\bigg)_{\rm Obs}+ 2\bigg(\frac{{\rm Q(CO_2)}}{{\rm Q(H_2O)}}\bigg)_{\rm Obs}\bigg]\,,
    \end{split}
\end{equation}
where the subscript ``Obs'' indicates the quantity when observed in the Solar System. This is analagous to the observed C/O ratio in Equation \ref{eq:c_to_ratio}. 

The goal of this subsection is to derive an approximate upper limit on the primordial C/O ratio from the observed ratio. In order to estimate this limit, we assume that the time averaged desorption of molecules in the interstellar medium affects all species equally (i.e., each species has a similar cross-section to galactic cosmic rays).  The ratio of production rates in the ISM likely varies at any given snapshot during this journey. However, we assume that the total time-averaged ratio of molecules desorbed is mediated by the thermodynamic properties of each species. As long as the comet is old enough such that this time averaging is a reasonable approximation, then the ratio of molecules desorbed is independent of the comet's age. In this idealized scenario, the composition of the processed comet surface reaches a steady state (depleted in the more volatile species) wherein the relative ablation rates of the various species are balanced by the addition of fresh unprocessed primordial material at the base of the active volume as the processed surface extends deeper into the body due to the ongoing erosion.  %We assume that the observed C/O ratio of the coma (i) represents the ISM processed surface of the nucleus and (ii)    provides a lower limit on the primordial C/O ratio of the nucleus. 

%Therefore, we assume that the ratio of the production rate, $Q(X)$, of a given species $X$ and ${\rm Q(H_2O)}$, remains constant throughout the processing in the ISM. This ratio is set by the thermodynamic properties of the species and the composition of the active area of the nucleus,

The steady-state composition of the processed surface is set by the thermodynamic properties of the species and the initial primordial composition of the comet. Assuming that the observed C/O ratio of the coma during its aparition in the solar system represents the composition of the ISM-processed surface of the nucleus,

\begin{equation}
    \bigg(\frac{{\rm Q(X)}}{{\rm Q(H_2O)}}\bigg)_{\rm Obs}=\frac{1}{\Phi_X}\,\bigg(\frac{{\rm Q(X)}}{{\rm Q(H_2O)}}\bigg)_{\rm Prim}\,,
\end{equation}
where the ratio $\Phi_X$ is defined as,
\begin{equation}\label{eq:Phi_X}
    \Phi_X=\,\bigg(\frac{\Delta  H(H_2O)/N_{A}+\gamma k\textit{T}_{\rm Sub}(H_2O)}{\Delta H(X)/N_{A}+\gamma k\textit{T}_{\rm Sub}(X)}\,\bigg)\,.
\end{equation}
In the case where the production rates of CO$_2$, CO and H$_2$O are measured, the primordial C/O ratio can be estimated as,
\begin{equation}\label{eq:ctoO_primordial}
\begin{split}
    \bigg({\rm C/O}\bigg)_{\rm Prim} =\\\bigg[\Phi_{\rm CO} \bigg(\frac{{\rm Q(CO)}}{{\rm Q(H_2O)}}\bigg)_{\rm Obs}+\Phi_{\rm CO_2} \bigg(\frac{{\rm Q(CO_2)}}{{\rm Q(H_2O)}}\bigg)_{\rm Obs}\bigg]\bigg/\\\bigg[ 1+ \Phi_{\rm CO}\bigg(\frac{{\rm Q(CO)}}{{\rm Q(H_2O)}}\bigg)_{\rm Obs}+ 2\,\Phi_{\rm CO_2}\bigg(\frac{{\rm Q(CO_2)}}{{\rm Q(H_2O)}}\bigg)_{\rm Obs}\bigg]\,.
    \end{split}
\end{equation}
The values of $\Phi_{\rm CO_2}=1.89$ and $\Phi_{\rm CO}=6.41$ are calculated using Equation \ref{eq:Phi_X} and the values in Table \ref{table:thermo}. Since $\Phi_{\rm CO_2},\Phi_{\rm CO}>1$, the measured C/O ratios of interstellar comets based on CO$_2$, CO and H$_2$O production rates measured in the Solar System are lower limits on the interstellar comets' primordial C/O ratios.   

\begin{figure}
\begin{center}
\includegraphics[scale=0.475,angle=0]{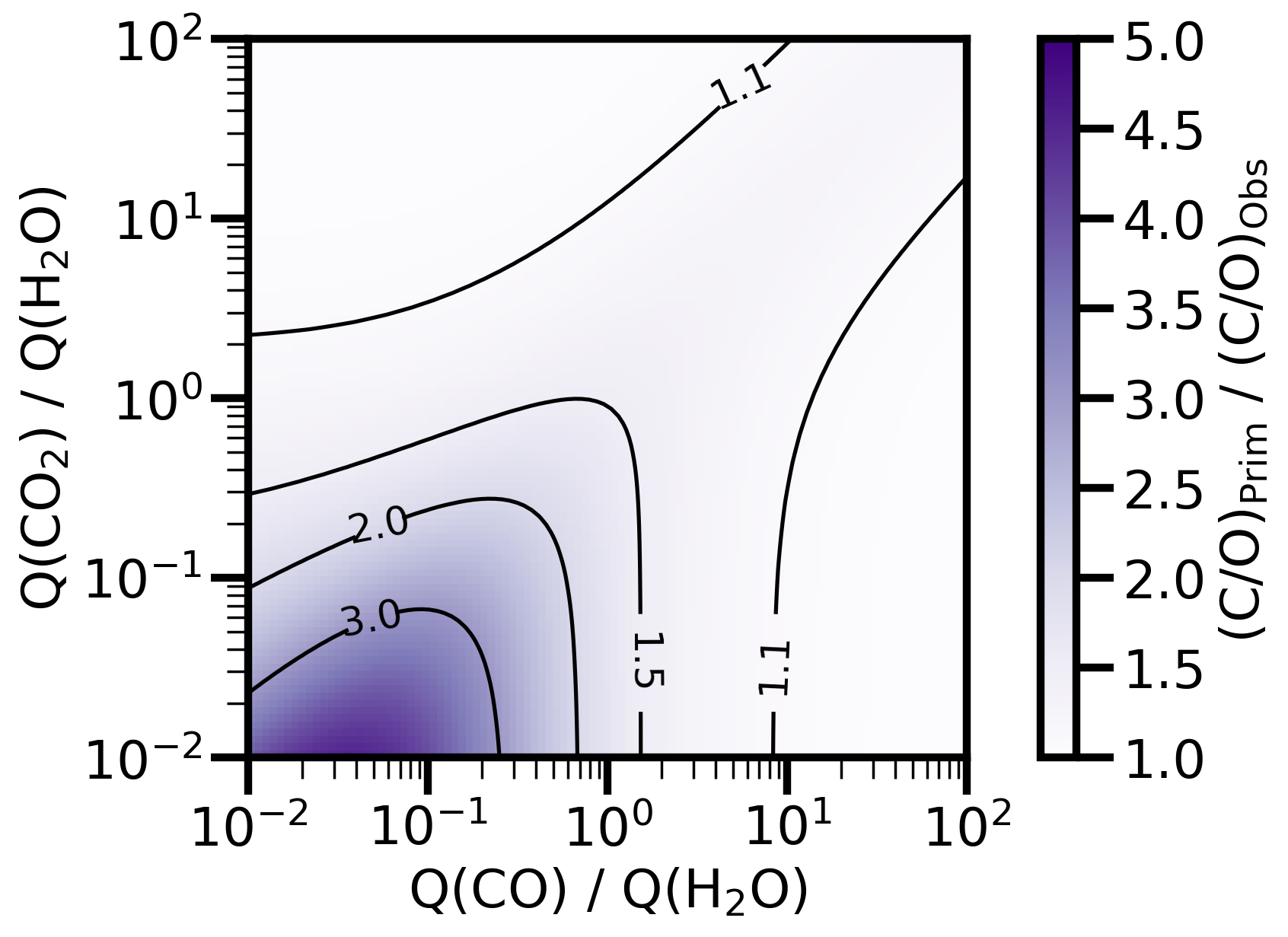}
\caption{The transformation from the observed to   primordial C/O ratio of an interstellar comet. The ratio is calculated by evaluating Equations \ref{eq:ctoO} and \ref{eq:ctoO_primordial}  for a range of  production rate ratios. Processing  decreases the observed C/O ratio due to  preferential desorption of CO and CO$_2$ relative to H$_2$O. }\label{Fig:primordial_transformation}
\end{center}
\end{figure}

In Figure \ref{Fig:primordial_transformation}, we show the ratio of the primordial to observed C/O ratios given by Equations \ref{eq:ctoO} and \ref{eq:ctoO_primordial} for a range of  production rates of CO and CO$_2$ relative to H$_2$O. This transformation can be applied to production rates of interstellar comets to estimate their primordial C/O values and can be extended to include measured production rates for any species. For Borisov, the production rates of CO and H$_2$O measured between 12/19/19 and 12/22/19 imply a C/O ratio $\sim0.56$, which would correspond to a primordial C/O ratio of $\sim0.89$. %However, we reiterate the point that though these assumptions are highly idealized, the observed C/O ratio of Borisov can confidently be seen as a lower limit on the primordial ratio. 

Admittedly,  the assumptions involved in transforming from the observed to primordial C/O ratio (Figure~\ref{Fig:primordial_transformation}) are highly idealized. The processing in the ISM of an interstellar comet is likely a much more complicated process than the one described here. Therefore, we only apply the transformation derived here to estimate limits on the primordial composition in \S \ref{sec:predictions}. The purpose of this calculation is \textit{not} to calculate a definitive primordial C/O ratio from the observed one. Interior compositional measurements obtained during an interception mission would provide a more accurate callibration of the transformation from observed to primordial compositional ratios. 

\subsection{Protective Shielding in the ISM}

Borisov was enriched in CO compared to most solar system comets, despite the fact that it had $\Delta {R}_{\rm SS}/\Delta {R}_{\rm ISM}<<1$. Moreover, if the non-gravitational acceleration of `Oumuamua was caused by the sublimation of CO \citep{Seligman2021}, then the C/O ratio would have been close to unity. Therefore, it may be that the preferential erosion of CO in an interstellar comet is a minor effect. 

It is possible that the  exposure to the galactic radiation environment produces a comet-like object with an insulating crust of volatile-depleted ``regolith'' of  low thermal conductivity \citep{Cooper2003}. This regolith could shield subsurface layers, leaving them intact. This was initially pointed out as a possible explanation for `Oumuamua's lack of coma by \citet{Jewitt2017} and elaborated upon with detailed thermal modelling by \citet{Fitzsimmons2017}.  \citet{Seligman2018}  demonstrated  that less than 10 cm of regolith coating was sufficent  to protect subsurface volatiles from sublimating, even when exposed to the solar radiation (Figure 1 in that paper). This layer is smaller than the layer removed from Borisov during its solar system passage estimated by \citet{Kim2020}. Such mantled regolith crusts could help to preserve the interior of an interstellar comet through interstellar space.

\begin{figure*}
\begin{center}
\includegraphics[scale=0.6,angle=0]{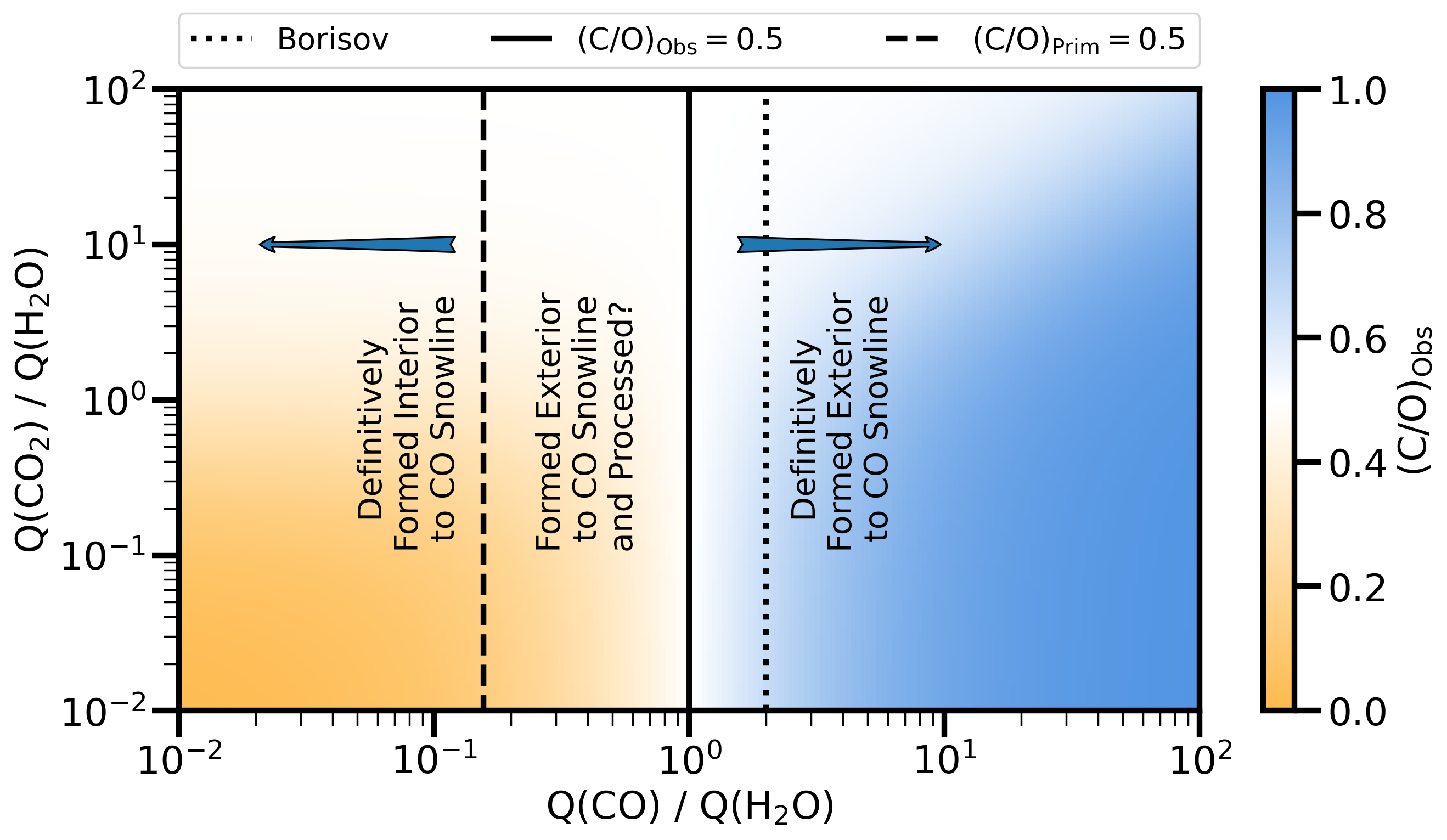}
\caption{The observed C/O ratio of an interstellar comet as a function of the measured production rates of CO and CO$_2$ relative to H$_2$O. The solid contour line indicates the region where the observed C/O ratio is 0.5. Since processing  preferentially removes CO and CO$_2$ relative to H$_2$O, comets to the right of this contour formed exterior to the CO snowline. The dashed line indicates where the primordial C/O ratio is 0.5, using the transformation presented in Figure \ref{Fig:primordial_transformation}. Objects to the left of this contour formed interior to the CO snowline. Objects in between both of these contours could have formed exterior to the CO snowline with primordial C/O ratios greater than 0.5, with their observed C/O ratios altered due to  processing. The location of Borisov is indicated with dotted lines, implying that it formed exterior to the CO snowline.  }\label{Fig:future_comets}
\end{center}
\end{figure*}

\section{Inferring The Formation Location of Future Interstellar Comets}\label{sec:predictions}

In this section, we investigate the feasibility of inferring the formation location of an interstellar comet relative to the CO snowline in its host protostellar disk.  Assuming that interstellar comet host stars have C/O ratios $\sim0.5$  (Section \ref{sec:stellar_CO}), a primordial composition with C/O$>0.5$ implies formation exterior to the CO snowline.

As we showed in the previous section,  processing increases the C/O ratio of an interstellar comet. However, mechanisms such as (i) self shielding (subsection 5.6), (ii) inefficient desorption, and (iii) sputtering could decrease the efficiency of this effect. Therefore,  measured C/O ratios can be interpreted as a  lower limit on the primordial ratio. In Figure \ref{Fig:future_comets}, we show the observed C/O ratio of an interstellar comet for a range of measured production rate ratios. In the overplotted contours, we show the regions where the observed and primordial C/O ratios are equal to 0.5, based on the transformations from processing.  A \textit{measured} C/O ratio $>0.5$  is definitive evidence that the interstellar comet formed exterior to the CO snowline in its host system. If the measured observed C/O ratio $<0.5$, but the primordial C/O ratio is $>0.5$, it is still possible that the comet formed exterior to the CO snowline and  experienced preferential desorption of CO during processing. If the measured C/O ratio $\le0.2$, then the object definitively formed interior to the CO snowline. 

In Figure \ref{Fig:future_comets}, we indicate the measured C/O ratio of Borisov, safely within the region that implies definitive formation exterior to the CO snowline. Future detections of interstellar comets and measurements of the production rates of CO$_2$, CO and H$_2$O will constrain the fraction of the population that formed exterior to the CO snowline.  

\section{Observational Capability and Planning}\label{sec:observations}

\subsection{Observatories and Science Objectives}
In this section, we describe  observations of an interstellar comet that would reveal its formation location and the facilities capable of obtaining them. We identify the primary science goal as constraining the elemental carbon to oxygen ratio in the gaseous coma of the interstellar comet and describe several other secondary objectives. The following quantities and ratios  would be  useful to measure for this goal:
\begin{itemize}
    \item Elemental ratios (C/O, N/O, S/O) (direct from UV, inferred from VIS/NIR/Sub-mm molecule abundances)
    \item Molecular abundances (NUV/VIS/NIR/Sub-mm)
    \begin{itemize}
        \item Molecules: H$_2$O, CO, CO$_2$ (N/UV, VIS, NIR, Sub-mm)
        \item Radicals: OH, CN, C$_2$, C$_3$ (NUV, VIS, NIR)
    \end{itemize}
\end{itemize}

 Production ratios of ``parent" volatiles like H$_2$O, CO$_2$, or CO at regular intervals during the apparition will provide a direct measurement of the C/O ratio. Large field-of-view observations of the broader atomic comae in the ultraviolet with fortuitous viewing geometry would  provide atomic carbon and oxygen abundances.  Regular observations of daughter molecules (produced via the breakdown of `parent" molecules) such as OH, CN, and C$_2$,  would offer an additional constraint on the bulk carbon-compound abundance. Measured daughter molecules would also enable comparisons with Solar System comets that do not have measured CO$_2$, CO and H$_2$O production rates.
 \setcounter{table}{3}
\begin{table*}
    \centering
    \begin{tabular}{c|c|c|c}
         & Wavelength &  Relevant & \\
         Science Goal & Regime & Observatories & Limitations  \\
         \hline
         \hline
         Atomic Abundances & 900-2100 \AA\ & HST (STIS, COS) & Small FOV, Operational lifetime\\
         (Direct) &  &  & operational lifetime \\
            & 3000-5000 \AA\ & VLT (UVES) &  \\
            & 4200-11000 \AA\ & Keck (HIRES) & \\
         & 1216 \AA\ & SOHO (SWAN) & Pointing, Limited to Hydrogen\\
         \hline
         Relative Atomic Abundances &  3200-10000 \AA & Keck &   \\
          (Inferred from Daughter Products)& 3000-25000 \AA\ & VLT (XSHOOTER) &  \\
          & 3200-10000 \AA\ & LBT & Monsoon Season Deadtime\\
          
         \hline
         Molecular Abundance & 1.0-5.0 $\mu{m}$ & Keck (NIRSPEC) & Target Brightness \\
         (H$_{2}$O, \textit{direct}) & & IRTF (iShell) & \\\hline
         Molecular Abundance & 2500-3400 \AA\ & Swift (UVOT) & Operational lifetime \\
         (H$_{2}$O, \textit{inferred}) & & & \\
         \hline
          Molecular Abundances (CO, CS) & 1400-3100 \AA\ & HST (STIS, COS) & Oversubscription  \\
         \hline
         Molecular Abundances & Sub millimeter& ALMA & February Maintenance Blackout  \\
         (H$_{2}$O, CO$_{2}$, CO) & Sub millimeter & SMT & Monsoon Season Deadtime \\
          & MIR & JWST &  ToO execution\\
    \end{tabular}
    \caption{Science goals, the observatories capable of making the relevant observations, and limiting factors to support comprehensive characterization of the next interstellar object.}
    \label{tab:STM}
\end{table*}

 \subsection{Facilities and Approaches}

 In this subsection, we identify observational facilities capable of achieving the identified science goals.
 For many of the goals, there are only a handful of ground or space-based facilities capable of achieving them. A non-exhaustive comparison of the objectives, required wavelength coverage, and relevant facilities are listed in Table \ref{tab:STM}.
 
 In general, measuring the C/O ratio directly  requires a census of all relevant ions, atoms, molecules, and radicals in the coma. However, a decent understanding can be achieved through monitoring the production rates of H$_2$O, CO, and CO$_2$ (Figure \ref{fig:ahearn_comets}). Measuring production rates of these species directly will require Hubble or JWST and a ground-based millimeter-wave facility such as  ALMA. CO is observable  from the ground at millimeter-wavelengths and in the NIR, while   H$_2$O and CO$_2$ require space-based infrared observations in most cases. With the  \textit{Spitzer} satellite  past its lifetime, JWST is the most capable space-based facility  sensitive to these infrared emissions. If the comet is sufficiently bright,    H$_2$O and CO can be measured with Keck-NIRSPEC or IRTF-iShell. The Stratospheric Observatory For Infrared Astronomy (SOFIA) \citep{Temi2018} could pursue similar objectives and measurements as JWST but with lessened sensitivity.
 
 The detection of CO in 2I/Borisov is an excellent example of a limiting detection for an interstellar comet \citep{Bodewits2020}. When Borisov was detected at  $\sim$2 au from both the Sun and the Earth, its V-band magnitude was $\sim17$. The CO detection required 17,901 integration seconds (5 orbits of exposures) with the HST COS instrument at that magnitude. While the relationship between the V-band magnitude and CO production rates is not well understood, future UV spectral observations should be planned assuming similar requirements. Although space-based UV or IR observations are the most reliable way to characterize these parent molecules, they also require the objects to be brighter than for proxy measurements. 
 
 Proxies for H$_2$O, like OH, OI, or atomic hydrogen can be obtained for dim comets under fortuitous conditions. For comets of lower apparent brightness, near-ultraviolet, visible, and near-infrared observations will remain viable for a significant fraction of the observable trajectory. The OH radical can be converted to the H$_2$O production rate (see \S \ref{sec:cometary_CO} and references therein) and is observable from the ground in the NUV. However, these observations require telescopes with instrumentation sensitivity at wavelengths comparable to the atmospheric cut-off. Moreover, the extinction of 3100 $\AA$ light is extremely sensitive to airmass, so the comet's sky position is  as important as its  brightness and H$_2$O production rate. To circumvent this, the H$_2$O production rate of Borisov was measured in the 0-0 band of OH emission with the Ultraviolet-Optical Telescope on the Swift Observatory \citep{Xing2020}. However, Swift is currently addressing a broken reaction wheel and not available for observations. OI emission  also approximates the H$_2$O production rate \citep{McKay2020}. This  requires  the oxygen emission to be sufficiently blue-or-red-shifted  from atmospheric emission by the geocentric motion of the comet. This is only feasible for a subset of comets with suitable trajectories. 

The carbon budget of an interstellar comet can be approximated by measuring production rates of carbon-bearing radicals CN, C$_2$, and C$_3$ between $\sim$3800 $\AA$ and $\sim$5600 $\AA$. This is obtainable with visible-wavelength spectroscopy or specialized filters like  the Hale-Bopp filter set \citep{2000Icar..147..180F} (which also encompasses an OH 0-0 transition). These specialized filter sets can be used to characterize faint comets not observable with visible spectroscopy. However,  few  specialized filter sets are  available outside of the Lowell Observatory and NOIRLab telescope networks. The adoption of the Hale-Bopp or a similar filter set at more sites -- especially very high altitude sites like Mauna Kea -- would greatly increase our ability to characterize faint interstellar comets.

Characterizing  daughter products will also permit  classification of interstellar comets according to the Solar System comet taxonomies (such as ``typical" vs. ``depleted," \citealt{Ahearn:1995}). Some comets exhibit apparently typical carbon budgets \citep{RaymondJC2022}  without easily detected emissions from these species \citep{2008AJ....136.2204S}. The prevalance of this effect in interstellar comets will also contextualize them within the populations native to the Solar System. 

 Generally, observations of carbon and oxygen atoms directly would require space-based ultraviolet observations, though exceptions do exist. HST's COS \citep{green_cosmic_2012} and STIS \citep{woodgate1998space} instruments are the only currently available space-based ultraviolet capable spectrographs sensitive to these emissions with guest observer capability. 

\subsection{Faint Interstellar Comets}

LSST will detect interstellar comets at or close to its limiting magnitude \citep{Hoover2021}. For these cases, some or all of the techniques outlined in the previous subsection cannot be applied. In this subsection, we highlight feasible alternative observational techniques when  spectroscopic observations cannot be obtained.

The secular light curve, or brightness variations as a function of  heliocentric distance, of a faint interstellar comet discovered by a survey such as the LSST should still be straightforward to obtain. A reliable measure of its brightness will be automatically obtained every few days by the discovery survey. Larger telescopes  will only be necessary when the comet is below the  detection limits of the discovery survey. Outbound observations of comets have precedent, provide  additional information about the evolution of activity, and contextualize the measurements obtained when the comets were brightest (such as Hale-Bopp, see \citealt{2012ApJ...761....8S}).

The secular light curve of a interstellar comet will  provide useful information regarding its composition. The  light curve of inactive objects is driven only by the heliocentric distance and the phase angle. The brightness variations of active objects  is dominated by dust lofted from the surface and fluorescence of gas molecules and radicals. These effects cause a steeper slope with respect to heliocentric distance  as the active volatile ratios change \citep{1997Sci...275.1915B}.  Brightness variations as an interstellar comet crosses various ice lines will  provide \textit{some} compositional constraints. For example, the extent to which the light curve is smooth or punctuated by large outbursts and periods of low activity at large heliocentric distances informs (i) which volatiles drive the activity and (ii) the compositional structure within the nucleus  \citep{2021PSJ.....2...48K}. For comets that are bright and active enough to appear extended in imaging observations, the morphology of cometary comae can also provide constraints on the overall activity state and  volatiles driving activity \citep{Kim2020}.

\subsection{Preparing for Observations of Future Interstellar Comets}

In this subsection, we outline an observational strategy for a future interstellar comet with the primary goal of measuring the C/O ratio. The fraction of comets ejected from exterior to the CO snowline will be further constrained with every interstellar comet characterized in this manner. The LSST should detect $\ge10$ interstellar comets over 10 years \citep{Hoover2021},  $\sim50\%$ of which have $b<b_{\rm 2I}$ (Figure \ref{Fig:cdf_v_b}) allowing for measurements of their C/O ratio. This  will provide a statistical sample of $\ge7$ objects to constrain the fraction ejected that formed exterior to the CO snowline, and yield insights into the efficiency of various scattering mechanisms.  

The  observations of Borisov serve as an example to build from for future interstellar comets. Multiple observations of the H$_{2}$O, CO$_{2}$, and CO as well as typical cometary radicals OH, CN, C$_{2}$ were obtained and revealed the object's unique characteristics. Ideally these observations  will be repeated with future comets. Additionally, observations at heliocentric distances beyond 3 au pre and post perihelia would inform the level to which ISM processing altered the surface composition. This would be vital information for estimating the primordial C/O ratio based on the measured one. Borisov was discovered  $\sim$3 months prior to perihelia, limiting time for pre-perihelion characterization of the molecular production rates. The drastic change in Borisov's H$_2$O production rate post-perihelion was explained by the removal of a CO-depleted and  H$_2$O enriched  surface  \citep{Bodewits2020}. Pre perihelia CO production rate measurements of a future interstellar comet would quantify the efficiency of this effect. Optimal observations of future interstellar comets would provide pre and post perihelia  characterization of the molecular abundances  on either side of the H$_2$O ice sublimation line at $\sim$3 au, for the extent to which discovery and observing geometry allows.  This corresponds to a minimum of 4 epochs with detections or upper limits on H$_{2}$O, CO$_{2}$, and CO as well as the more typical cometary radicals OH, CN, and C$_{2}$.

 The  apparent magnitude of an interstellar comet is agnostic to the extended brightness of the coma.  However, the activity level controls the brightness in exposures where the bandpass is sensitive to fluorescence of the sublimating gas or reflected sunlight from lofted dust. The apparition of  2I/Borisov was notably poor, partly because the comet only reached a geocentric distance of $\sim2$au with a poor solar elongation angle. An LPC or JFC with the same absolute magnitude as Borisov would have  been too poor of a target propose observations for. However, due to its extra-solar origin, it was well characterized due to a global campaign. Therefore, limitations on the detectable activity level   with certain telescopes for solar system comets should not be applied to interstellar comets. 
 
The observatories capable of performing this suite of observations are presented in Table \ref{tab:STM}. The instrumentation and observatories are capable of measuring the C/O ratio of an interstellar comet independently. However, this system is susceptible to single-point failures. For example, in the event that the HST is unavailable during an interstellar comet apparition due to operational limitations, H, C, O and S atomic abundances cannot be  measured directly with a single observation. If an interstellar comet is at high declination in the northern hemisphere during the monsoon season in Arizona and Swift UVOT is unavailable, it can only be characterizble by the IRTF and Keck observatories. This limits our ability to perform independent measurements of atomic, molecular, and radical abundances to ensure accuracy.  Therefore, accurately measuring  C/O ratios of interstellar comets is directly linked to the stability of space-based observing platforms.

\section{Conclusions}\label{sec:conclusions}
In this paper, we advocated for obtaining production rate measurements of CO$_2$, CO and H$_2$O of future interstellar comets. These measurements will provide lower limits on their primordial carbon to oxygen ratios.  In \S \ref{sec:CO_tracer}, we described how this ratio traces  the  formation location relative to the CO snowline. This technique is already used for extrasolar planets  \citep{Oberg2011} .  In \S \ref{sec:stellar_CO}, we reviewed  current measurements of stellar C/O ratios. Since the scatter in stellar C/O ratios is low, the C/O ratio of an interstellar comet is a reasonable tracer for formation location within a protostellar disk. 

In \S \ref{sec:cometary_CO}, we reviewed measurements of the C/O ratio in Solar System comets. We showed that measurements of CO$_2$, CO and H$_2$O best approximate the C/O ratio, given their high abundance relative to other species. These measurements have revealed that most  Solar System comets  formed interior to the CO snowline. Similar measurements of interstellar comets will constrain the fraction of ejected comets that  formed exterior to the CO snowline.  2I/Borisov, C/2016 R2, and possibly `Oumuamua likely formed in this region.

In \S \ref{sec:processing}, we quantified the relative importance of processing in the interstellar medium and the Solar System as a function of the lifetime and trajectory for an interstellar comet. We concluded that volatile production rates are unlikely to be representative of the primordial composition for most objects that will be detected with the LSST. Because of preferential desorption of CO and CO$_2$ relative to H$_2$O in the ISM, measured C/O ratios are lower limits on the primordial one. In \S \ref{sec:predictions}, we show that production rate ratios of ${\rm Q}({\rm CO})/{\rm Q}({\rm H_2O})<.2$ and ${\rm Q}({\rm CO})/{\rm Q}({\rm H_2O})>1$  indicate formation interior and exterior to the CO snowline,  respectively. It is  possible that the primordial composition of an interstellar comet can be measured during a disintegration event, a tidal disruption event,  or in-situ  during an impactor rendezvous mission. Additionally, measurements of the H$_2$O production rates of interstellar comets may encode information regarding the galactic star formation history \citep{Lintott2022}.

In \S \ref{sec:observations},  we reviewed the relevant observations of atoms, molecules, and radicals that constrain the C/O ratio of an interstellar comet. A range of space and ground based observatories covering the UV to sub-mm wavelengths will be required to measure abundances independently and characterize the composition accurately. Comprehensive atomic and molecular measurements would ideally be attempted on either side of the H$_2$O  sublimation point ($\sim 3$au) prior to and post perihelia for a minimum of four observation epochs. We argued that compositional constraints can be obtained for faint interstellar comets that do not permit detailed spectroscopic characterization by monitoring for activity variation as a function of heliocentric distance via imaging campaigns.

Knowledge of the formation location of the population of ejected interstellar comets will yield key insights into the mechanisms driving planetary formation and evolution in exoplanetary systems and the Solar System. If the population of interstellar comets mostly consists of objects that formed exterior to the CO snowline, then a natural interpretation is that the Solar System also produced a population of CO-enriched comets that were ejected via early dynamical instability. It appears feasible that 2I/Borisov, and potentially R2 and 'Oumuamua, may be repesentative of this distinct class of comets. Moreover, the age of these interstellar comets could provide us with an estimate for the relative timing of dynamical instabilities in exoplanetary systems. Based on theoretical modelling of the early timing of the instability in the Solar System \citep{Grav2011,Buie2015,Nesvorny2018b,Clement2018,Clement2019,deSousa2019,Nesvorny2021,Morgan2021}, and the recent discovery of an excess of free-floating planets in the Upper Scorpius $<10$ MYR stellar association \citep{Miret-Roig2022}, it seems feasible that the majority of  interstellar comets are ejected within the first $ <10$ MYR of a planetary system's lifetime. If the majority of the population of interstellar comets formed interior to the CO snowline, this would imply that early giant planet migration is common in this region.

\section{Acknowledgements}
We thank Dave Jewitt, Konstantin Batygin, Jack Palmer, Juliette Becker, Kaitlin Kratter, Greg Laughlin, Fred Adams and Robert Jedicke for useful conversations. We thank the scientific editor, Maria Womack, and the two anonymous reviewers for insightful comments and constructive suggestions that strengthened the scientific content of this manuscript. 

ADF acknowledges support from the National Science Foundation Graduate Research Fellowship Program under Grant No. DGE-1746045.  MRK acknowledges support from the Australian Research Council through its \textit{Future Fellowships} scheme, award FT180100375. LAR gratefully acknowledges support from the Research Corporation for Science Advancement through a Cottrell Scholar Award.
MM acknowledges by NASA through the NASA Hubble Fellowship grant HST-HF2-51485.001-A awarded by the Space Telescope Science Institute. KEM acknowledges support from NASA through Rosetta Data Analysis Program (RDAP) grant 80NSSC19K1306.

\bibliography{sample63}{}
\bibliographystyle{aasjournal}

\clearpage

\begin{landscape}
\setcounter{table}{0}
\begin{longrotatetable}
\movetabledown=2.2in
%\begin{deluxetable*}{@{}c|c|c|c|c|c|c|c|c|c@{}}
\begin{deluxetable*}{@{}l |c| c| c| c | c | c | c | c |c | c r@{}}
\tablecaption{ Measured production rate ratios of CO$_2$ and CO with respect to H$_2$O, heliocentric distances and inferred C/O ratios of solar system comets.   For comets with multiple production rates measurements, we quote the mean ratios and range of distances. We only include comets that have measured production rates of H$_2$O and CO and/or CO$_2$. Reported upper limits are multiplied by a factor of $1/3$ in the calculation of the C/O ratio. For multiple reported observations within 1 day of each other, we report the mean of the measurements and uncertainties. Exceptions: Nightly observations of CO and H$_2$O were reported for Comet C/2002 T7 LINEAR over a 5 day timespan, for which we quote only the reported weighted average. Most of the data was drawn from Tables 1 and 2 in \citet{DelloRusso2016}, Table 1 in \citet{Ahearn2012} and Tables 1 and 3 in \citet{Ootsubo2012}. For C/2016 R2 we report the mean values spanning observations through January and February of 2018 as reported by \citet{McKay2019}. Updated measurements for 103P are from HST observations and the EPOXI mission flyby \citep{ahearn2011}.}
\label{tab:production rates}
\tablewidth{\linewidth}
\tabletypesize{\scriptsize}
\tablehead{
\colhead{Comet} & \colhead{Date} & \colhead{r$_H$ [au]} & \colhead{CO/H$_2$O} & \colhead{(CO/H$_2$O)$_\textrm{err}$} & \colhead{CO$_2$/H$_2$O} & \colhead{(CO$_2$/H$_2$O)$_\textrm{err}$} & \colhead{C/O} & \colhead{(C/O)$_\textrm{err}$} & \colhead{Reference} }
\startdata
%\\
1P/Halley & 3/13/1986 & 0.90 & 0.035 & 0.006 & 0.035 & 0.006 &  &   &  {\citet{Bockelee2004}} \\
 &  3/10/1986 & 0.86 & 0.065 & 0.006 & 0.059 & 0.005 &  &   &  {\citet{Feldman1997}}  \\
 &  3/11/1986 & 0.87 & 0.043 & 0.006 & 0.051 & 0.005 &  &   &  {\citet{Feldman1997}}  \\
 &  3/16/1986 & 0.95 & 0.082 & 0.006 & 0.063 & 0.009 &  &   &  {\citet{Feldman1997}}  \\
 &  3/18/1986 & 0.97 & 0.041 & 0.006 & 0.028 & 0.01 &  &   &  {\citet{Feldman1997}}  \\
 &  &  &  &  &   &  &   &   &  \\
Mean &  &0.91 &0.05 &0.01 &0.05& 0.01 &  0.09& 0.01  &  \\\hline
153P/Ikeya-Zhang & 4/13/02 & 0.78 & 0.048 & 0.009 &  &   &   &   &  CO:\citet{Disanti2002} \\
& & & & & & & & & H$_2$O:\citet{delloRusso2002} \\
 &  3/20/02 & 0.51 & 0.033 & 0.007 &  &   &   &   &  CO:\citet{Biver2006} \\
& & & & & & & & & H$_2$O:\citet{delloRusso2002} \\
 &  05/12/02 & 1.26 & 0.043 & 0.005 &  &   &   &   &  CO:\citet{Biver2006} \\
& & & & & & & & & H$_2$O:\citet{Lecacheux2003}\\
Mean &  &0.85 &0.04 &0.01 &0.00& 0.00 &  0.04& 0.01  &  \\\hline
8P/Tuttle & 1/27/08 & 1.03 & 0.0045 & 0.001 &  &   &   &   &  {\citet{Bohnhardt2008}} \\
Mean &  &1.03 &0.004 &0.001 &0.000& 0.000 &  0.004& 0.001  &  \\\hline
64P/Swift-Gehrels & 11/23/09 & 2.27 & $<0.02$ &  &  0.2905 & 0.017 &  &   &  {\citet{Ootsubo2012}} \\
Mean &  &2.27 &0.007 &0.000 &0.290& 0.017 &  0.187& 0.011  &  \\\hline
19P/Borrelly & 12/30/08 & 2.19 & $<0.24$ &  &  0.2410 & 0.009 &  &   &  {\citet{Ootsubo2012}} \\
Mean &  &2.19 &0.080 &0.000 &0.241& 0.009 &  0.206& 0.006  &  \\\hline
103P/Hartley 2 & 09/16/91 & 0.96 &  &   &  0.041 &  &   &   &  {\citet{Weaver1994}} \\
 &  12/31/97 & 1.08 &  &   &  0.097 & 0.016 &  &   &  {\citet{Crovisier1999}}  \\
 &  11/04/10 & 0.96 & 0.0015–0.0045 &  &  $<0.2$ &  &   &   &  {\citet{Weaver2011}}\\
Mean &  &1.00 &0.003 &0.000 &0.068& 0.016 &  0.063& 0.014  &  \\\hline
144P/Kushida & 04/18/09 & 1.70 & 0.014 & 0.0001 & 0.1590 & 0.002 &  &   &  {\citet{Ootsubo2012}} \\
Mean &  &1.70 &0.014 &0.0001 &0.159& 0.002 &  0.130& 0.002  &  \\\hline
67P/Churyumov-Gerasimenko & 11/2/08 & 1.84 & $<0.2$ &  &  0.0700 & 0.003 &  &   &  {\citet{Ootsubo2012}} \\
Mean &  &1.84 &0.067 &0.0000 &0.070& 0.030 &  0.113& 0.025  &  \\\hline
73P/S-W 3C & 5/3/06 & 1.07 & $<$0.0026 &  &   &   &   &   &  \citet{DelloRusso2007}  \\
 &  5/27/06 & 0.95 & 0.0047 & 0.0019 &  &   &   &   &  {\citet{DiSanti2007}}  \\
 &  5/30/06 & 0.95 & 0.0058 & 0.0018 &  &   &   &   &  {\citet{DiSanti2007}} \\
Mean &  &0.99 &0.004 &0.0019 &0.000& 0.000 &  0.004& 0.002  &  \\\hline
73P/S-W 3B & 5/9/06 & 1.03 & $<$0.0019 &  &   &   &   &   &  \citet{DelloRusso2007} \\
Mean &  &1.03 &0.001 &0.0000 &0.000& 0.000 &  0.001& 0.0005  &  \\\hline
157P/Tritton & 12/30/09 & 1.48 & $<$0.14 &  &  0.0945 & 0.004 &  &   &  {\citet{Ootsubo2012}} \\
Mean &  &1.48 &0.047 &0.0000 &0.095& 0.004 &  0.114& 0.003  &  \\\hline
22P/Kopff & 04/22/09 & 1.61 & $<$0.03 &  &  0.2 & 0.02 &  &   &  {\citet{Ootsubo2012}}  \\
 &  12/11/09 & 2.43 &$<$0.21 &  &  0.074 & 0.007 &  &   &  {\citet{Ootsubo2012}} \\
Mean &  &2.02 &0.040 &0.0000 &0.137& 0.013 &  0.135& 0.010  &  \\\hline
81P/Wild 2 & 12/14/09 & 1.74 &$<$0.03 &  &  0.15 & 0.015 &  &   &  {\citet{Ootsubo2012}} \\
Mean &  &1.74 &0.010 &0.0000 &0.150& 0.015 &  0.122& 0.011  &  \\\hline
88P/Howell & 07/03/09 & 1.74 &$<$0.06 &  &  0.25 & 0.02 &  &   &  {\citet{Ootsubo2012}} \\
Mean &  &1.74 &0.020 &0.0000 &0.250& 0.002 &  0.178& 0.001  &  \\\hline
118P/Shoemaker-Levy 4 & 09/08/09 & 2.18 &$<$0.21 &  &  0.23 & 0.023 &  &   &  {\citet{Ootsubo2012}} \\
Mean &  &2.18 &0.070 &0.0000 &0.230& 0.023 &  0.196& 0.015  &  \\\hline
9P/Tempel 1 & 7/3/05 & 1.49 & 0.1087 & 0.049 & 0.0696 & 0.02 &  &   &  {\citet{Feaga2007}}  \\
 &  7/5/05 & 1.51 & 0.043 & 0.012 &  &   &   &   &  {\citet{Mumma2005}} \\
Mean &  &1.50 &0.076 &0.0305 &0.070& 0.020 &  0.120& 0.030  &  \\\hline
116P/Wild 4 & 05/15/09 & 2.22 &$<$0.17 &  &  0.08 & 0.008 &  &   &  {\citet{Ootsubo2012}} \\
Mean &  &2.22 &0.057 &0.0000 &0.080& 0.008 &  0.112& 0.007  &  \\\hline
C/1979 Y1 Bradfield & 01/10/1980 & 0.71 & 0.035 & 0.004 & 0.035 & 0.004 &  &   &  {\citet{Feldman1997}} \\
Mean &  &0.71 &0.035 &0.0040 &0.035& 0.004 &  0.063& 0.005  &  \\\hline
C/1989 X1 Austin & 05/09/1990 & 0.83 & 0.017 & 0.008 & 0.021 & 0.008 &  &   &  {\citet{Feldman1997}} \\
Mean &  &0.83 &0.017 &0.0080 &0.021& 0.008 &  0.036& 0.011  &  \\\hline
C/1990 K1 Levy & 08/26/1990 & 1.38 & 0.041 & 0.008 & 0.069 & 0.008 &  &   &  {\citet{Feldman1997}}  \\
 &  09/18/1990 & 1.13 & 0.084 & 0.015 & 0.133 & 0.015 &  &   &  {\citet{Feldman1997}} \\
Mean &  &1.25 &0.062 &0.0115 &0.101& 0.011 &  0.129& 0.013  &  \\\hline
C/1995 O1 Hale-Bopp & 01/21/97 & 1.49 & 0.267 & 0.0029 &  &   &   &   &  CO:{\citet{DiSanti2001}} \\
& & & & & & & & & H$_2$O:{\citet{DelloRusso2000}} \\
 &  02/23/97 & 1.11 & 0.241 & 0.0016 &  &   &   &   &  CO:{\citet{DiSanti2001}} \\
& & & & & & & & & H$_2$O:{\citet{DelloRusso2000}} \\
 &  03/01/97 & 1.06 & 0.271 & 0.0011 &  &   &   &   &  CO:{\citet{DiSanti2001}} \\
& & & & & & & & & H$_2$O:{\citet{DelloRusso2000}} \\
 &  04/09/97 & 0.93 & 0.276 & 0.0022 &  &   &   &   &  CO:{\citet{DiSanti2001}} \\
& & & & & & & & & H$_2$O:{\citet{DelloRusso2000}} \\
 &  04/16/97 & 0.95 & 0.297 & 0.0028 &  &   &   &   &  CO:{\citet{DiSanti2001}} \\
& & & & & & & & & H$_2$O:{\citet{DelloRusso2000}} \\
 &  04/30/97 & 1.05 & 0.222 & 0.0023 &  &   &   &   &  CO:{\citet{DiSanti2001}} \\
& & & & & & & & & H$_2$O:{\citet{DelloRusso2000}} \\
 &  05/01/97 & 1.06 & 0.280 & 0.0024 &  &   &   &   &  CO:{\citet{DiSanti2001}} \\
& & & & & & & & & H$_2$O:{\citet{DelloRusso2000}} \\
Mean &  &1.09 &0.265 &0.0022 &0.000& 0.000 &  0.209& 0.002  &  \\\hline
C/1996 B2 Hyakutake & 03/15/1996 & 1.24 & 0.149 & 0.033 &  &   &   &   &  CO:{\citet{Biver1999}} \\
& & & & & & & & & H$_2$O:{\citet{Gerard1998}} \\
 &  04/01/1996 & 0.89 & 0.178 & 0.047 &  &   &   &   &  CO:{\citet{Biver1999}} \\
& & & & & & & & & H$_2$O:{\citet{Gerard1998}} \\
 &  04/10/1996 & 0.67 & 0.328 & 0.02 &  &   &   &   &  CO:{\citet{Biver1999}} \\
& & & & & & & & & H$_2$O:{\citet{Bertaux1998}}\\
Mean &  &0.93 &0.218 &0.0333 &0.000& 0.000 &  0.179& 0.027  &  \\\hline
C/1999 H1 Lee & 08/24/1999 & 1.12 & 0.04 & 0.01 &  &   &   &   &  {\citet{Biver2000}} \\
 &  08/20/1999 & 1.06 & 0.018 & 0.002 &  &   &   &   &  {\citet{Mumma2001}}\\
Mean &  &1.09 &0.029 &0.0060 &0.000& 0.000 &  0.028& 0.006  &  \\\hline
C/1999 S4 LINEAR & 07/4/2000 & 0.87 & 0.009 & 0.003 &  &   &   &   &  {\citet{Mumma2001linear}}  \\
 &  07/13/2000 & 0.81 & 0.004 & 0.003 &  &   &   &   &  {\citet{Mumma2001linear}} \\
Mean &  &0.84 &0.006 &0.0030 &0.000& 0.000 &  0.006& 0.003  &  \\\hline
C/1999 T1 McNaught-Hartley & 01/06/2001 & 1.23 & 0.224 & 0.1 &  &   &   &   &  {\citet{Biver2006}}  \\
 &  01/30/2001 & 1.41 & 0.102 & 0.1 &  &   &   &   &  {\citet{Biver2006}}  \\
 &  02/05/2001 & 1.44 & 0.127 & 0.1 &  &   &   &   &  {\citet{Biver2006}} \\
Mean &  &1.36 &0.151 &0.1000 &0.000& 0.000 &  0.131& 0.087  &  \\\hline
C/2000 WM1 LINEAR & 11/23/2001 & 1.32 & 0.0052 & 0.001 &  &   &   &   &  {\citet{Radeva2010}} \\
Mean &  &1.32 &0.005 &0.0010 &0.000& 0.000 &  0.005& 0.001  &  \\\hline
C/2001 A2 LINEAR & 07/10/2001 & 1.17 & 0.0386 & 0.011 &  &   &   &   &  CO: \citet{Magee2008} \\
& & & & & & & & & H$_2$O:\citet{Gibb2007}\\
Mean &  &1.17 &0.039 &0.0110 &0.000& 0.000 &  0.037& 0.011  &  \\\hline
C/2002 T7 LINEAR (DN) & 05/03-09/2004 & 0.69 & 0.019 & 0.003 &  &   &   &   &  CO/H$_2$O: \citet{Anderson2010} \\
& & & & & & & & & H$_2$O from OH:\citet{DiSanti2006}\\
Mean &  &0.69 &0.019 &0.0030 &0.000& 0.000 &  0.019& 0.003  &  \\\hline
C/2004 Q2 Machholz & 11/29/04 & 1.48 & 0.0507 & 0.007 &  &   &   &   &  {\citet{Bonev2009}} \\
Mean &  &1.48 &0.051 &0.0070 &0.000& 0.000 &  0.048& 0.007  &  \\\hline
C/2006 OF2 Broughton & 9/16/08 & 2.43 & $<0.04$ &  &  0.235 & 0.03 &  &   &  {\citet{Ootsubo2012}}  \\
 &  3/28/09 & 3.20 &$<$0.26 &  &  0.58 & 0.06 &  &   &  {\citet{Ootsubo2012}} \\
Mean &  &2.82 &0.050 &0.0000 &0.407& 0.045 &  0.245& 0.024  &  \\\hline
C/2006 M4 SWAN (DN) & 11/7/06 & 1.08 & 0.0049 & 0.0022 &  &   &   &   &  {\citet{DiSanti2009}}  \\
 &  11/9/06 & 1.10 & 0.0051 & 0.0021 &  &   &   &   &  {\citet{DiSanti2009}} \\
Mean &  &1.09 &0.005 &0.0022 &0.000& 0.000 &  0.005& 0.002  &  \\\hline
C/2007 W1 Boattini (DN) & 7/10/08 & 0.90 & 0.045 & 0.0051 &  &   &   &   &  {\citet{Villanueva2011}} \\
Mean &  &0.90 &0.045 &0.0051 &0.000& 0.000 &  0.043& 0.005  &  \\\hline
C/2007 N3 Lulin &  &  1.26-1.70 & 0.0223 & 0.002 & 0.1095 & 0.001 &  &   &  {\citet{Ootsubo2012}} \\
 &  2/1/09 & 1.26 & .0219 & 0.001 &  &   &   &   &  {\citet{Gibb2012}}\\
Mean &  &1.48 &0.022 &0.0015 &0.110& 0.001 &  0.106& 0.001  &  \\\hline
C/2007 Q3 Siding Spring & 03/03/09 & 3.29 & $<0.1$ &  &  0.19 & 0.07 &  &   &  {\citet{Ootsubo2012}} \\
Mean &  &3.29 &0.033 &0.0000 &0.190& 0.070 &  0.158& 0.050  &  \\\hline
C/2008 Q3 Garradd & 07/05/09 & 1.81 & 0.26 & 0.03 & 0.28 & 0.03 &  &   &  {\citet{Ootsubo2012}}  \\
 &  07/06/09 & 1.81 & 0.22 & 0.03 & 0.25 & 0.03 &  &   &  {\citet{Ootsubo2012}}  \\
 &  01/03/10 & 2.96 &$<$0.56 &  &  0.64 & 0.06 &  &   &  {\citet{Ootsubo2012}} \\
Mean &  &2.19 &0.222 &0.0300 &0.390& 0.040 &  0.306& 0.025  &  \\\hline
C/2009 P1 Garradd & 9/18/11 & 2.02 & 0.135 & 0.015 &  &   &   &   &  {\citet{Paganini2012}}  \\
 &  9/21/11 & 2.00 & 0.116 & 0.012 &  &   &   &   &  {\citet{Paganini2012}}  \\
 &  9/21/11 & 2.01 & 0.046 & 0.011 & 0.116 & 0.009 &  &   &  {\citet{McKay2015}}  \\
 &  10/10/2011 & 1.85 & 0.062 & 0.011 & 0.117 & 0.009 &  &   &  {\citet{McKay2015}}  \\
 &  1/25/2012 & 1.62 & 0.147 & 0.032 &  &   &   &   &  {\citet{McKay2015}}  \\
 &  2/27/2012 & 1.69 & 0.195 & 0.05 & 0.056 & 0.009 &  &   &  {\citet{McKay2015}}  \\
 &  3/26/2012 & 2.00 & 0.63 & 0.32 & 0.085 & 0.02 &  &   &  {\citet{Feaga2014}}  \\
& & & & & & & & & H$_2$O:\citet{Bodewits_2014} \\
Mean &  &1.88 &0.190 &0.0644 &0.094& 0.012 &  0.206& 0.048  &  \\\hline
29P Schwassmann-Wachmann 1 & 11/18/09 & 6.18 & 4.64 & 0.4 & $<0.04$ &  &   &   &  {\citet{Ootsubo2012}} \\
Mean &  &6.18 &4.640 &0.4000 &0.013& 0.000 &  0.821& 0.071  &  \\\hline
C/2012 F6 Lemmon & 3/31/13 & 0.75 & 0.0425 & 0.0064 &  &   &   &   &  \citet{Paganini2014}  \\
 &  4/1/13 & 0.75 & 0.0382 & 0.0063 &  &   &   &   &  \citet{Paganini2014} \\
Mean &  &0.75 &0.040 &0.0063 &0.000& 0.000 &  0.039& 0.006  &  \\\hline
C/2012 S1 ISON (DN) & 11/15/13 & 0.59 & 0.0151 & 0.0017 &  &   &   &   &  {\citet{disanti2016}}  \\
 &  11/16/13 & 0.56 & 0.0091 & 0.0027 &  &   &   &   &  {\citet{disanti2016}}  \\
 &  11/17/13 & 0.53 & 0.0134 & 0.0011 &  &   &   &   &  {\citet{disanti2016}}  \\
 &  11/22/13 & 0.35 & 0.0181 & 0.0037 &  &   &   &   &  {\citet{disanti2016}} \\
Mean &  &0.51 &0.014 &0.0023 &0.000& 0.000 &  0.014& 0.002  &  \\\hline
C/2010 G2 Hill & 1/10/12 & 2.5 & 0.91 & 0.23 &  &   &   &   &  \citet{Kawakita2014} \\
Mean &  &2.50 &0.910 &0.2300 &0.000& 0.000 &  0.476& 0.120  &  \\\hline
2P/Encke & 11/5/03 & 1.19 & $<.0177$ &  &   &   &   &   &  \citet{Radeva2013} \\
Mean &  &1.19 &0.006 &0.0000 &0.000& 0.000 &  0.006& 0.000  &  \\\hline
%6P/d'Arrest &  &  $$ &  &   &  0.023 &  &   &   &   \\
%17P/Holmes &  &  $$ &  &   &  0.069 &  &   &   &   \\
C/2006 W3 Christensen & 12/21/08 & 3.66 & 3.61 & 0.3 & 1.02 & 0.1 &  &   &  \citet{Ootsubo2012}  \\
 &  06/16/09 & 3.13 & 0.98 & 0.06 & 0.42 & 0.06 &  &   &  \citet{Ootsubo2012} \\
Mean &  &3.40 &2.295 &0.1800 &0.720& 0.080 &  0.637& 0.042  &  \\\hline
C/2006 Q1 (McNaught) & 06/03/08 & 2.78 & $<0.1$ &  &  0.45 & 0.05 &  &   &  \citet{Ootsubo2012}  \\
 &  02/23/09 & 3.64 &$<$0.47 &  &  0.44 & 0.05 &  &   &  \citet{Ootsubo2012} \\
Mean &  &3.21 &0.095 &0.0000 &0.445& 0.050 &  0.272& 0.025  &  \\\hline
C/2007 G1 (LINEAR) & 08/20/08 & 2.80 & $<0.17$ &  &  0.23 & 0.02 &  &   &  \citet{Ootsubo2012} \\
Mean &  &2.80 &0.057 &0.0000 &0.230& 0.020 &  0.189& 0.013  &  \\\hline
21P/Giacobini-Zinner & 10/2/98 & 1.25 & 0.1 & 0.06 &  &   &   &   &  \citet{Mumma2000} \\
 &  10/26/98 & 1.10 &$<$0.032 &  &   &   &   &   &  {\citet{Weaver1999}} \\
Mean &  &1.18 &0.055 &0.0000 &0.000& 0.000 &  0.052& 0.000  &  \\\hline
C/2013 R1 Lovejoy & 10/24/13 & 1.34 & 0.099 & 0.02 &  &   &   &   &  \citet{Paganini2014b}\\
Mean &  &1.34 &0.099 &0.0200 &0.000& 0.000 &  0.090& 0.018  &  \\\hline
C/2016 R2 & 01-02/18 & 2.8 & 312.5 & 40.0 & 56.9 & 40.0 &  &   &  \citet{McKay2019}  \\
Mean &  &2.80 &312.500 &40.0000 &56.900& 50.000 &  0.864& 0.150  &  \\
\enddata
\end{deluxetable*}
\end{longrotatetable}

\end{landscape}
\clearpage
\begin{landscape}
\setcounter{table}{1}

\begin{longrotatetable}
\movetabledown=2.2in
\begin{deluxetable*}{@{}c|c|c|c|c|c|c|c|c|c|c@{}}
\tablecaption{Various production rates for molecules observed for 2I/Borisov. Productions rates are in units of $10^{24}$ molecules/second, except for H$_2$O and CO which are 10$^{26}$. Errors are included where possible. }  \label{tab:borisov}
\tablewidth{\linewidth}
\tabletypesize{\scriptsize}
\tablehead{
\colhead{Date} & \colhead{r$_H$ [au]} & \colhead{${\rm Q({CN})}$} & \colhead{${\rm Q({C_2})}$} & \colhead{${\rm Q({C_3})}$} & \colhead{${\rm Q({OH})}$ } & \colhead{ ${\rm Q({NH_2})}$} & \colhead{${\rm Q({H_2O})}$} & \colhead{${\rm Q({CO})}$} & \colhead{${\rm Q({HCN})}$}& \colhead{Reference} }
\startdata
9/20/19&2.67&$3.7\pm0.4$&$<4$&& & &&&&\citet{Fitzsimmons:2019}\\\hline
9/20/19&2.67&$<5$&$<8$&& & &&&&\citet{Kareta:2019}\\\hline
9/27/19&2.56&&&& & &$<8.2$&&&\citet{Xing2020} \\\hline
10/1/19&2.50&$1.1\pm2.0$&$<2.5$&& & &&&&\citet{Kareta:2019}\\\hline
10/1/19&2.51&$1.8\pm0.1$&$<0.9$&$<0.3$& & &&&&\citet{Opitom:2019-borisov}\\\hline
10/2/19&2.50&$1.9\pm0.1$&$<0.6$&$<0.2$&$<20$& &&&&\citet{Opitom:2019-borisov}\\\hline
10/9/19&2.41&$1.59\pm0.09$&$<0.44$&& & &&&&\citet{Kareta:2019}\\\hline
10/10/19&2.39&$1.69\pm0.04$&$<0.162$&& & &&&&\citet{Kareta:2019}\\\hline
10/11/19&2.38&&&& & &$6.3\pm1.5$&&&\citet{McKay2020} \\\hline
10/13/19&2.36&$2.1\pm0.1$&$<0.6$&$<0.3$&$<20$& &&&&\citet{Opitom:2019-borisov}\\\hline
10/18/19&2.31&$1.9\pm0.6$&&& & &&&&\citet{Opitom:2019-borisov}\\\hline
10/20/19&2.29&$1.6\pm0.5$&&& & &&&&\citet{Opitom:2019-borisov}\\\hline
10/26/19&2.23&$1.9\pm0.3$&&& & &&&&\citet{Kareta:2019}\\\hline
10/31/19&2.18&$2.0\pm0.2$&&& & &&&&\citet{lin2020}\\\hline
11/1/19&2.17&&&& & &$7.0\pm1.5$&&&\citet{Xing2020}\\\hline
11/4/19&2.15&$2.4\pm0.2$&$5.5\pm0.4$&$0.03\pm0.01$& & &&&&\citet{lin2020}\\\hline
11/10/19&2.12&$1.9\pm0.5$&&& & &&&&\citet{Bannister2020}\\\hline
11/14/19&2.09&$1.8\pm0.2$&$1.1$&& &$4.20$&&&&\citet{Bannister2020}\\\hline
11/17/19&2.08&$1.9\pm0.5$&&& & &&&&\citet{Bannister2020}\\\hline
10/28/19-11/18/19&2.21-2.06& &&& & &&&$<6.3$&\citet{Bergman2022}\\\hline
11/25/19&2.04&$1.6\pm0.5$&&& & &&&&\citet{Bannister2020}\\\hline
11/26/19&2.04&$1.8\pm0.2$&&& & &&&&\citet{Bannister2020}\\\hline
11/26/19&2.04&$1.5\pm0.5$&$1.1$&& &$4.80$&&&&\citet{Bannister2020}\\\hline
11/30/19&2.01&$3.36\pm0.25$&$1.82\pm0.6$&$0.197\pm0.052$& & &&&&\citet{Aravind2021}\\\hline
12/1/19&2.01&&&& & &$10.7\pm1.2$&&&\citet{Xing2020}\\\hline
12/3/19&2.01&&&& & &&$3.3\pm0.8$&&\citet{yang2021}\\\hline
12/11/19&2.01&&&& & &&$7.5\pm2.3$&&\citet{Bodewits2020}\\\hline
12/15-16/2019&2.02&&&& & &&$4.4\pm0.7$&$0.7\pm0.11$&\citet{Cordiner2020}\\\hline
12/19-22/2019&2.03&&&& & &$4.9\pm0.9$&$6.4\pm1.4$&&\citet{Bodewits2020}\\\hline
12/21/19&2.03&&&& & &$4.9\pm0.9$&&&\citet{Xing2020}\\\hline
12/22/19&2.03&$6.68\pm0.27$&$2.3\pm0.82$&$0.714\pm0.074$& & &&&&\citet{Aravind2021}\\\hline
12/30/19&2.07&&&& & &&$10.7\pm6.4$&&\citet{Bodewits2020}\\\hline
1/13/20&2.16&&&& & &$<5.6$&$8.7\pm3.1$&&\citet{Bodewits2020} \\\hline
1/14/20&2.17&&&& & &$<6.2$&&&\citet{Xing2020}\\\hline
2/17/20&$2.54$&&&&&&$<2.3$&&&\citet{Xing2020}\\\hline
\enddata
\end{deluxetable*}
\end{longrotatetable}

\end{landscape}
\clearpage
\end{document}